%% file: root.tex
\documentclass[a4paper, 10pt, conference]{ieeeconf}      % Use this line for a4 paper

\IEEEoverridecommandlockouts                              % This command is only needed if 
\usepackage{amsmath} % assumes amsmath package installed
\usepackage{amssymb}  % assumes amsmath package installed
\usepackage{cite}
\usepackage{graphicx}
\usepackage[caption=false, font=footnotesize]{subfig}

\usepackage{tikz}
\usepackage{pgfplots}
\usetikzlibrary{external}

\usepackage{algorithm}
\usepackage{algorithmic}

\newcommand\copyrighttext{%
	\footnotesize Copyright $\copyright$ 2019 IEEE.
	Personal use of this material is permitted.
	Permission from IEEE must be obtained for all other uses, in any current or future media, including reprinting/republishing this material for advertising or promotional purposes, creating new collective works, for resale or redistribution to servers or lists, or reuse of any copyrighted component of this work in other works.  DOI:~10.1109/CAVS.2019.8887769 }%
\newcommand\copyrightnotice{%
	\begin{tikzpicture}[remember picture,overlay]%
	\node[anchor=south,yshift=10pt] at (current page.south) {\fbox{\parbox{\dimexpr\textwidth-2cm}{\copyrighttext}}};%
	\end{tikzpicture}%
	\vspace{-10pt}%
}

\title{\LARGE \bf
Subjective Logic-based Identification of Markov Chains and Its Application to CAV's Safety$^\ast$
}

\author{Johannes M{\"u}ller$^{1 \dagger}$, Thomas Griebel$^{1 \dagger}$, Michael Gabb$^{2}$, and Michael Buchholz$^{1}$% <-this % stops a space
	\thanks{*Part of this work was financially supported by the Federal Ministry of Economic Affairs and Energy of Germany within the program "Highly and Fully Automated Driving in Demanding Driving Situations" (project MEC-View, grant number 19A16010I).}% <-this % stops a space	
	\thanks{$\dagger$ J. M{\"u}ller and T. Griebel are both first authors with equal contribution.}	
	\thanks{$^{1}$Johannes M{\"u}ller, Thomas Griebel, and Michael Buchholz are with the Institute of Measurement, Control and Microtechnology,
		Ulm University, D-89081 Ulm, Germany
		{\tt\small \{johannes-christian.mueller, thomas.griebel, michael.buchholz\}@uni-ulm.de}
	}%
	\thanks{$^{2}$Michael Gabb is with the Robert Bosch GmbH, 74003 Heilbronn, Germany
		{\tt\small michael.gabb@de.bosch.com}}%
}

\begin{document}

\maketitle
\copyrightnotice
\thispagestyle{empty}
\pagestyle{empty}

%%%%%%%%%%%%%%%%%%%%%%%%%%%%%%%%%%%%%%%%%%%%%%%%%%%%%%%%%%%%%%%%%%%%%%%%%%%%%%%%
\begin{abstract}

A reliable estimation of the communication channel which connects automated vehicles is an important step towards the safety of connected and automated vehicles. The communication channel is usually modeled as Markov chain with slowly time-varying transition rates and is identified through statistics on the observed state transitions. However, the classical identification approach lacks a measure on how reliable the identification results, and thus the channel, are. In this work, we propose an identification method, based on the subjective logic theory, which features such a reliability measure in terms of statistical uncertainty. We demonstrate through simulations that the proposed method is capable of quickly responding to parameter changes. Furthermore, it is shown that the transition rates of the Markov chain are tracked with high accuracy. Finally, we validate our results by a real-world experiment. 

\end{abstract}

%%%%%%%%%%%%%%%%%%%%%%%%%%%%%%%%%%%%%%%%%%%%%%%%%%%%%%%%%%%%%%%%%%%%%%%%%%%%%%%%
\section{INTRODUCTION}

Estimating the reliability of cooperative information, which is shared among automated vehicles (AVs), is a big step on the way towards evolving AVs into connected and automated vehicles (CAVs). Reliable cooperative information, arriving with low latency at the CAV, can significantly increase the efficiency and benefit the safety of automated driving \cite{Siegel2018}. However, if the restrictive requirements of safety critical applications, such as motion planning of a CAV, to the latency are not met, this may result in harm \cite{Siegel2018}.

Identifying the properties of a communication channel to estimate its reliability is a long studied problem dating back to the works of Gilbert \cite{gilbert1960capacity} and Elliot \cite{elliott1963estimates} in the 1960s, which resulted in the Gilbert-Elliot model. Although more sophisticated models exist today, the Gilbert-Elliot model is still frequently used \cite{Iacovazzi2017}. A channel model as well as the identified parameters thereof are usually used subsequently to estimate the reliability of the communication link \cite{Chakravorty2017}, e.g., for vehicle-to-anything (V2X) applications \cite{Zhou2019}. However, existing approaches neglect the statistical uncertainty corresponding to the identified results. We, in turn, argue that these statistical uncertainties can significantly contribute to the reliability estimation and thus should not be neglected.

In this work, we propose a method to identify the communication channel which is modeled as Markov chain and is used for V2X communication based on subjective logic (SL). SL is a mathematical theory that allows to explicitly model the statistical uncertainty of the identified probabilistic model \cite{Joesang2016}. Thus, in addition to the identified parameters, the reliability of the identification result is estimated. This information, on the one hand, can be used in the CAVs to decide whether a CAV should rely on the cooperative information or fall back to AV mode. On the other hand, we show that the SL result can be used to efficiently recognize and track parameter changes such as jumps or drifts.

The contribution of this paper is twofold: from a theoretical perspective, to the best of our knowledge, there has not yet been developed a method where time-varying Markov chains are integrated into the SL framework. From a practical perspective, we show how the proposed method can be used to estimate the reliability of the V2X communication link of CAVs. If the communication link is not reliable enough, the CAV can terminate the link and is able to fall back to AV mode. Thus, the safety of CAVs is increased. 

\section{RELATED WORK} \label{sec:RelatedWork}

From a theoretical perspective, to the best of our knowledge, evidence-based subjective logic (EBSL) \cite{vskoric2016flow} is the extension of SL most closely related to our proposition. The key idea of EBSL is to combine flow-based reputation systems with the uncertainty concept of SL to determine indirect computational trust through a trust network. Flow-based reputation systems, in turn, have their mathematical foundation in Markov chains. Thus, EBSL indirectly links Markov chains to SL. However, EBSL requires the underlying Markov chain to be static. Furthermore, the underlying Markov chain is constrained by having only two states, as EBSL is only defined for binary opinions (see Section \ref{SL_basics}). In contrast, in this work, we propose an algorithm to identify Markov chains with an arbitrary finite number of states. In addition, the proposed algorithm is able to track the parameters of a time-varying Markov chain.

From a practical perspective, reliability estimation with respect to quality of service (QoS) has to be distinguished from the semantic reliability estimation. For example, in \cite{muller2019subjective}, we have proposed such a semantic reliability estimation mechanism and demonstrated its ability through experiment. In contrast, in this work, we focus on the reliability estimation of the QoS.
While several works, e.g., \cite{Iacovazzi2017}, \cite{Chakravorty2017}, \cite{Zhou2019}, have modeled the communication link and thereby achieved good results for highly reliable communication links, however, none of these works considers the statistical uncertainty within the identified model parameters used for the reliability estimation. 

\section{SUBJECTIVE LOGIC BASICS} \label{SL_basics}
In this section, we briefly summarize the subjective logic basics used in this paper. 
The definitions and theorems are based on \cite{Joesang2016}, where further details can be found.

\textbf{Definition 1 (Subjective Logic Opinion):} 
\textit{Let $X \in \mathbb{X}$ be a random variable of the finite domain $\mathbb{X}$ with the cardinality $k = |\mathbb{X}| \geq 2$. 
A subjective logic opinion (opinion in short) is an ordered triple $\omega_X = (\boldsymbol{b}_X,u_X,\boldsymbol{a}_X)$ with
\begin{subequations}
	\small
	\begin{align}
		\boldsymbol{a}_X : \mathbb{X} \mapsto [0,1], \qquad 1 &= \sum\limits_{x \in \mathbb{X}} \boldsymbol{a}_X (x) \, ,\\
		\boldsymbol{b}_X : \mathbb{X} \mapsto [0,1], \qquad 1 &= u_X + \sum\limits_{x \in \mathbb{X}} \boldsymbol{b}_X (x)\, .
	\end{align}
\end{subequations}
Hereby, $\boldsymbol{b}_X$ is the belief mass distribution over $\mathbb{X}$, $u_X$ is the uncertainty mass representing the lack of evidence and $\boldsymbol{a}_X$ is the base rate distribution over $\mathbb{X}$ representing the prior probability.
Moreover, the projected probability
\begin{equation}
\boldsymbol{P}_X(x) = \boldsymbol{b}_X(x) + \boldsymbol{a}_X(x) u_X, \quad \forall x \in \mathbb{X}
\end{equation}  
represents the expected outcome of an opinion.}

Subjective logic opinions express some information about a random variable $X$ in terms of belief, uncertainty, and base rate.
One key aspect of opinions is that they are linked to classical probability theory by a bijective mapping to Beta or Dirichlet probability density functions (PDF).

\textbf{Definition 2 (Dirichlet Distribution):} 
\textit{Let $\mathbb{X}$ be a domain of $W$ mutually disjoint values, $\boldsymbol{r}_X$ be the evidence for outcome $x \in \mathbb{X}$ with $\boldsymbol{r}_X (x) \geq 0$ $\forall x \in \mathbb{X}$, $\boldsymbol{a}_X$ a prior distribution over $\mathbb{X}$, and $\boldsymbol{p}_X$ the probability distribution over $\mathbb{X}$. Then, the PDF
\begin{align}
	\small
	\text{Dir}_X(\boldsymbol{p}_X,\boldsymbol{r}_X,\boldsymbol{a}_X) = &\frac{\Gamma \left( \sum \limits_{x \in \mathbb{X}} (\boldsymbol{r}_X (x) + \boldsymbol{a}_X (x) W) \right)}{\prod \limits_{x \in \mathbb{X}} \Gamma ( \boldsymbol{r}_X (x) + \boldsymbol{a}_X (x) W ) } \cdot \nonumber \\  
	&\prod \limits_{x \in \mathbb{X} } \boldsymbol{p}_X (x) ^{\boldsymbol{r}_X (x) + \boldsymbol{a}_X (x) W - 1} \, ,
	\label{eq:Dirichlet}
\end{align} 
where $\boldsymbol{r}_X (x) + \boldsymbol{a}_X (x) W \geq 0$ and $\boldsymbol{p}_X (x) > 0$ for $\boldsymbol{r}_X (x) + \boldsymbol{a}_X (x) W < 1$, is called Dirichlet PDF. A Dirichlet PDF with $W=2$ is called $\beta$-distribution. In (\ref{eq:Dirichlet}), $\Gamma( \, \cdot \, )$ is the well-known Gamma function \cite{Bronshtein2007}.}

\textbf{Theorem 1 (Equivalent Mapping \cite{Joesang2016}):} \textit{Let $\omega_X = (\boldsymbol{b}_X,u_X,\boldsymbol{a}_X)$ be a subjective logic opinion and $\text{Dir}_X(\boldsymbol{p}_X,\boldsymbol{r}_X,\boldsymbol{a}_X)$ a Dirichlet PDF over the same variable $X$ on domain $\mathbb{X}$. Then, the equivalent mapping
\begin{equation}
\label{eq:Mapping}
\small
\! \, \left . \begin{matrix}
\boldsymbol{b}_X(x) &\!\!\!\!\!= \frac{\boldsymbol{r}_X(x)}{W + \sum \limits_{x_i \in \mathbb{X}} \boldsymbol{r}_X(x_i)} \\
\\
u_X &\!\!\!\!\!= \frac{W}{W + \sum \limits_{x_i \in \mathbb{X}} \boldsymbol{r}_X(x_i)} 
\end{matrix} \right \} \!\!\! \iff \!\!\!
\left \{ \begin{array}{ll}
\boldsymbol{r}_X(x) = \frac{ W \boldsymbol{b}_X(x)}{u_X} \\
\\
1 = u_X + \sum \limits_{x_i \in \mathbb{X}} \boldsymbol{b}_X(x_i)
\end{array} \right .  \!\!\!\!,	
\end{equation}
transforms the Dirichlet distribution into the subjective logic opinion and vice versa.}

To combine opinions from various sources, there exists multiple fusion operators to merge these opinions. 

\textbf{Definition 3 (Aleatory Cumulative Belief Fusion):} \textit{Let $\omega_X^A$ and $\omega_X^B$ be source A and B's respective opinions over the same variable $X$ on domain $\mathbb{X}$. Let $\omega_X^{(A \diamond B)}$ be the opinion such that
\begin{equation}
	\small
	\omega_X^{(A \diamond B)} = \left \{  \begin{array}{ll}
		\boldsymbol{b}_X^{(A \diamond B)} &= \frac{ \boldsymbol{b}_X^{A}(x) u_X^{B} + \boldsymbol{b}_X^{B}(x) u_X^{A} }{ u_X^{A} + u_X^{B} - u_X^{A} u_X^{B} } \\
		\\[-6pt]
		u_X^{(A \diamond B)} &= \frac{ u_X^{A} u_X^{B} }{ u_X^{A} + u_X^{B} - u_X^{A} u_X^{B} }\\
		\\[-6pt]
		\boldsymbol{a}_X^{(A \diamond B)} &= \frac{ \boldsymbol{a}_X^{A}(x) u_X^{B} + \boldsymbol{a}_X^{B}(x) u_X^{A} }{ u_X^{A} + u_X^{B} - 2 u_X^{A} u_X^{B} } 
		\\[6pt]
		&\quad - \frac{ (\boldsymbol{a}_X^{A} (x) + \boldsymbol{a}_X^{B}(x)) u_X^{A} u_X^{B} }{ u_X^{A} + u_X^{B} - 2 u_X^{A} u_X^{B} } 	 	
	\end{array}\right . \, ,
\end{equation}
where $0 < u_X^A < 1$ and $0 < u_X^B < 1$, then the operator $\oplus$ in $\omega_X^{A \diamond B} = \omega_X^A \oplus \omega_X^B$
is called aleatory cumulative belief fusion.}

To obtain trust or belief from transitive trust paths, trust discounting is often used; for further details refer to \cite{Joesang2016}. 

\textbf{Definition 4 (Trust Discounting):} \textit{Let $\omega_X^A$ be source A's opinions over the variable $X$ on domain $\mathbb{X}$. Furthermore, let $\boldsymbol{p}_d$ be the discount probability.
Then, the function $\mathcal{D}\{\omega_X^A, \, \boldsymbol{p}_d\}$ that yields the opinion $\omega_X^{\text{disc}}$ such that
\begin{equation}
	\label{eq:TrustDiscounting}	
	\small
	\omega^{\text{disc}} = \mathcal{D}\{\omega_X^A, \, \boldsymbol{p}_d\} = \left \{ \begin{array}{ll}
		\boldsymbol{b}^{\text{disc}}(x) &\!\!\!\!\!= \boldsymbol{p}_{d} \; \boldsymbol{b}_X^{A}(x) \\
		u^{\text{disc}} &\!\!\!\!\!= 1 - \boldsymbol{p}_{d} \sum \limits_{x \in \mathbb{X}} \!  \boldsymbol{b}_X^{A}(x) \\
		\boldsymbol{a}^{\text{disc}}(x) &\!\!\!\!\!= \boldsymbol{a}_X^{A}(x) 
	\end{array}\right . 
\end{equation}
holds, is called trust discounting.}

\section{SUBJECTIVE LOGIC-BASED IDENTIFICATION OF A MARKOV CHAIN}
In this section, starting from the problem formulation, an overview of the SL-based identification algorithm is given and the respective steps are explained in detail.
\subsection{Problem Formulation} \label{sec:ProblemFormulation}
Let $\{X_{t_m} | t_m > 0, m \in \mathbb{N}\}$ be a Markov chain \cite{Bronshtein2007} with random variables $X_{t_m}$ at time $t_m$ and state space $Z = \{1, \ldots, N\}$. Then, the transition probabilities are defined as $P (X_{t_{m+1}}=j| X_{t_{m}}=i)=p_{ij}(t_m,t_{m+1})$ such that the so-called transition matrix is given by $\boldsymbol{P}(t_m,t_{m+1}) = \big[ p_{ij}(t_m,t_{m+1})\big]_{1\leq i,j \leq N}$. Using $\boldsymbol{P}$, the probability vector  $\boldsymbol{x}_{t_{m+1}} \in [0,1]^{N}$ can be calculated with
\begin{equation}
\label{eq:ProblemFormulation}
\boldsymbol{x}_{t_{m+1}} = \boldsymbol{P}(	t_{m}, 	t_{m+1}) \cdot 	\boldsymbol{x}_{t_{m}} \; .
\end{equation} 
The goal is to identify  $\boldsymbol{P}(t_{m},t_{m+1})$ from realizations. Further, a reliability measure is needed that indicates how certain the identified $\boldsymbol{P}(t_{m},t_{m+1})$ is based on statistical evidence.

\subsection{Algorithm} \label{sec:Overview}
The key idea of the algorithm is, firstly, to form a rough opinion of the transition matrix based on a buffered sliding window of samples and, consequently, check the opinion for consistency with the previous opinion on that matrix. If the opinions concur, the new opinion will be merged with the previous one, which results in an overall opinion that is based on more statistical data and, thus, features less statistical uncertainty. If, in turn, the opinions do not concur, the previous opinion will be discarded and the new one is chosen. Hence, the algorithm can react quickly to sudden changes such as jumps in the parameters. In contrast, slow changes in the parameters, e.g., drifts, are compensated for, as the new data updates the opinion and the old data is weighted less due to trust discounting. Algorithm \ref{alg:Algorithm} gives an overview of one iteration of the algorithm, while the two basic steps, namely the preprocessing and the consistency check, are discussed in detail in the following sections.
\begin{algorithm}
	\caption{SL-based Identification of Markov Chains}
	\label{alg:Algorithm}
	\small
	\begin{algorithmic}[1]
		\renewcommand{\algorithmicrequire}{\textbf{Input:}}
		\renewcommand{\algorithmicensure}{\textbf{Output:}}
		\REQUIRE Number of states $N \in \mathbb{N}$, statistical prior $\boldsymbol{A} = \left[ \boldsymbol{a}_i^T \right]_{1 \leq i \leq N}$ %\boldsymbol{a}_1, \ldots, \boldsymbol{a}_N \right)$
		with $\boldsymbol{a}_i \in [0,1]^{N}$,
		incoming stream of observations $\{o_k\}_{k \in \mathbb{N}}$, time $t_m, t_{m+1} > 0$, window length
		$l_w \in \mathbb{N}$, $W \in \mathbb{N}$, previous result $\boldsymbol{\Omega}^{t_{m}}$, discount probabilities $\boldsymbol{p}_d , \tilde{\boldsymbol{p}}_d  \in [0,1]^{N}$
		\ENSURE Estimated transition matrix $\boldsymbol{P}(t_{m},t_{m+1}) \in [0,1]^{N \times N}$ ,
		corresponding opinion $\boldsymbol{\Omega}^{t_{m+1}}$
		\vspace{5pt}
				
		\STATE $ \mathcal{W}^{t_{m+1}} \gets$ BufferWindow$\left(\, \{o_k\}_{m \, l_w \leq k \leq (m+1) \, l_w}\right)$
		
		\STATE $ s_{ij}^{t_{m+1}} \gets$ TransitionStatistics$(\, \mathcal{W}^{t_{m+1}} \,)$
		
		\STATE $\tilde{b}_{ij}^{t_{m+1}} = \frac{s_{ij}^{t_{m+1}}}{ \left( \sum_{l=1, l \neq j}^{N} s_{il}^{t_{m+1}} \right) + W}$
		
		\STATE $\tilde{u}_{i}^{t_{m+1}} = \frac{W}{ \left( \sum_{l=1}^{N} s_{il}^{t_{m+1}} \right) + W}$
		
		\STATE $\tilde{\omega}_{i}^{t_{m+1}} = (\tilde{\boldsymbol{b}}_i^{t_{m+1}}, \, \tilde{u}_{i}^{t_{m+1}}, \boldsymbol{a}_i)$
		\vspace{5pt}
		\STATE TrustDiscount$\left( \tilde{b}_{ij}^{t_{m+1}}, \, b_{ij}^{t_m}, \, \boldsymbol{p}_d, \,   \tilde{\boldsymbol{p}}_d \right)$
		\IF {isConsistent$(\, \tilde{b}_{ij}^{t_{m+1}}, \, b_{ij}^{t_m} \,)$ == TRUE }
		\STATE ${\omega}_{i}^{t_{m+1}} \gets \tilde{\omega}_{i}^{t_{m+1}} \oplus {\omega}_{i}^{t_{m}}$
		\ELSE
		\STATE ${\omega}_{i}^{t_{m+1}} \gets \tilde{\omega}_{i}^{t_{m+1}}$
		\ENDIF
		
		\STATE $\boldsymbol{P}(t_{m}, t_{m+1}) \gets \left[ \boldsymbol{b}_{i}^{t_{m+1}} + u_i^{t_{m+1}} \cdot \boldsymbol{a}_{i} \right]_{1 \leq i \leq N}$
		
		\STATE $ \boldsymbol{\Omega}^{t_{m+1}} \gets \left[ \omega_i^{t_{m+1}} \right]_{1 \leq i \leq N} $
		
		\RETURN  $\boldsymbol{P}(t_{m},t_{m+1})$, $\boldsymbol{\Omega}^{t_{m+1}}$
	\end{algorithmic} 
\end{algorithm}

\subsection{Preprocessing} \label{sec:Preprocessing}
In order to reach some statistical reliability, $l_w$ samples are buffered in the sliding window $\mathcal{W}^{t_{m+1}}$ where $l_w$ is the window length. The choice of $l_w$ is a trade-off. On the one hand, large values of $l_w$ result in high statistical reliability, thus, the consistency check has to deal with less statistical uncertainty and the system does not immediately react to some statistical variations in the data. On the other hand, small window sizes result in higher dynamics. Hence, the CAV is able to react more quickly to parameter changes.

Then, the statistics $s_{ij}^{t_{m+1}}$ are determined on the individual transitions from state $i$ to $j$ observed in $\mathcal{W}^{t_{m+1}}$. Based on these statistics, the respective opinions 
\begin{equation}
	\label{eq:WindowOpinions}
	\tilde{\omega}_{i}^{t_{m+1}} = (\tilde{\boldsymbol{b}}_i^{t_{m+1}}, \tilde{u}_{i}^{t_{m+1}}, \boldsymbol{a}_i) \; , \quad  \; 1 \leq i \leq N
\end{equation}

are calculated using equivalent mapping \eqref{eq:Mapping}. In \eqref{eq:WindowOpinions}, $\boldsymbol{a}_i$ is the a priori known statistical prior, $\tilde{u}_{i}^{t_{m+1}}$ is the uncertainty and $\tilde{\boldsymbol{b}}_i^{t_{m+1}}$ is the belief vector for the state $i$ at time $t_{m+1}$.

\subsection{Consistency Test} \label{sec:ConsistencyTest}
During the preprocessing, the opinion $\tilde{\omega}_{i}^{t_{m+1}}$ is calculated under the assumption that the actual transition matrix remains constant within the observation window $\mathcal{W}^{t_{m+1}}$. This assumption, however, is not always true, as the parameters might have changed within $\mathcal{W}^{t_{m+1}}$ with the probability $\boldsymbol{p}_d$. 
Even more likely, with the probability  $\tilde{\boldsymbol{p}}_d$, the actual transition matrix might have changed since $t_m$. 
This is accounted for by applying trust discounting on both, $\omega_{i}^{t_m} \gets \mathcal{D}\{ \omega_{i}^{t_m}, \, \boldsymbol{p}_d \}$ and $\tilde{\omega}_{i}^{t_{m+1}} \gets \mathcal{D}\{ \tilde{\omega}_{i}^{t_{m+1}}, \, \tilde{\boldsymbol{p}}_d \} $.

To check whether or not the transition matrix, estimated in $\tilde{\omega}_{i}^{t_{m+1}}$, is still consistent with the previously determined opinion $\omega_{i}^{t_m}$, the opinions are compared with respect to their \textit{degree of conflict (DC)} \cite{Joesang2016}. Hereby, the \textit{DC} is defined as
\begin{align}
	\label{eq:DC}
	\small
	\textit{DC}_i (t_m,t_{m+1}) = &\frac{1}{2} \sum_{j=1}^{N} | p_{ij} (t_{m-1}, t_m) - \tilde{p}_{ij} (t_{m}, t_{m+1}) | \cdot \nonumber\\ &(1 - u_i^{t_m}) (1 - \tilde{u}_i^{t_{m+1}})\; ,  \; 1 \leq i \leq N .
\end{align} 

If the opinions $\omega_{i}^{t_m}$ and $\tilde{\omega}_{i}^{t_{m+1}}$ are similar, the \textit{DC} is expected to be small, but not zero due to statistical variations. Thus, a threshold $\theta > 0$ is introduced for the $DC$. If $\textit{DC} \leq \theta$, the opinions are considered to concur, thus $\omega_{i}^{t_{m+1}}$ is updated using cumulative belief fusion, i.e., ${\omega}_{i}^{t_{m+1}} \gets \tilde{\omega}_{i}^{t_{m+1}} \oplus {\omega}_{i}^{t_{m}}$.
In contrast, if $\textit{DC} > \theta$, the opinions are considered to be inconsistent which indicates that the transition matrix has significantly changed between $t_m$ and $t_{m+1}$. Thus, the previous opinion $\omega_{i}^{t_m}$ is discarded and replaced by $\tilde{\omega}_{i}^{t_{m+1}}$. 
Finally, the results are summarized in matrix notation, i.e., $\boldsymbol{P}(t_{m}, t_{m+1})$ and $\boldsymbol{\Omega}^{t_{m+1}}$.

\section{SIMULATIONS} \label{sec:Simulation}
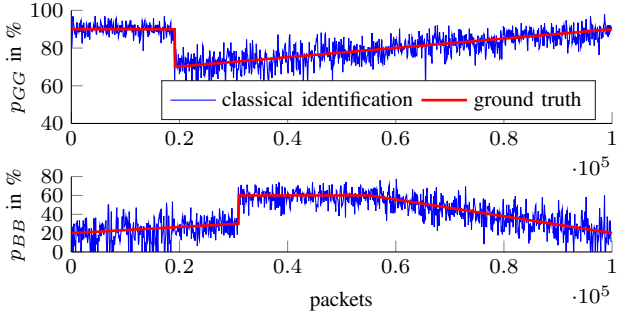
\begin{figure}[bthp]
	\centering
	\begin{tikzpicture}
	\footnotesize
	\input{img/ClassicalIdentificationSim.tex}
	\end{tikzpicture}
	\caption{Results of the classical approach for the simulated LTE channel.}
	\label{fig:ClassicalIdentificationSim}
\end{figure}

\begin{figure}
	\subfloat[Identification results.]{
		\begin{tikzpicture}
			\footnotesize
			\input{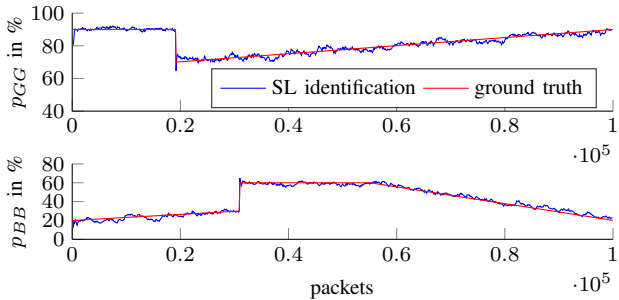}
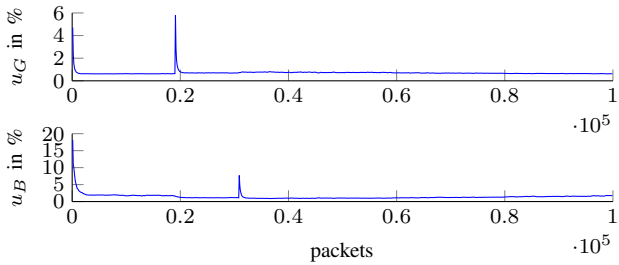
		\end{tikzpicture}
	} \\
	\subfloat[Uncertainty.]{
		\begin{tikzpicture}
			\footnotesize
			\input{img/SimUncetainty.tex}
		\end{tikzpicture}
	}
	\caption{Results of the SL-based algorithm for the simulated LTE channel. }
	\label{fig:SimIdentificationSL}
\end{figure}

The identification algorithm is exemplarily evaluated through simulations of a communication channel with burst errors. The channel is modeled as Markov chain with two states: \textit{good} (G) and \textit{bad} (B). Thus, (\ref{eq:ProblemFormulation}) turns into
\begin{equation}
	\label{eq:TwoStateMarkovChain}
	\small
	\boldsymbol{x}_{t_{m+1}} = \begin{bmatrix}
		p_{GG}(t_{m},t_{m+1}) & \!\!\! 1 - p_{GG}(t_{m},t_{m+1}) \\
		1-p_{BB}(t_{m},t_{m+1}) & \!\!\! p_{BB}(t_{m},t_{m+1})
	\end{bmatrix} \cdot \boldsymbol{x}_{t_{m}} .
\end{equation}
Fig.~\ref{fig:ClassicalIdentificationSim} shows the ground truth of an exemplary scenario as well as the result of a classical identification of (\ref{eq:TwoStateMarkovChain}) using data windows of length $l_w = 100$. For this scenario, a Long Term Evolution (LTE) radio link is assumed to be evaluated for packet loss before the Hybrid Automated Response Request (HARQ) \cite{Sauter2014}. 
Typically, to maximize the effective throughput, the radio link is configured such that about $10 \, \%$ of the packets have to be corrected by the HARQ mechanism \cite{Sauter2014}. As a result, usually, less than $1 \, \%$ of the packets need further corrections.
 Hence, to simulate a realistic scenario, we start with $p_{GG} = 90 \, \%$. The scenario is evaluated over $100\,000$ packets, while at packet $19\,081$ and $30\,851$ sudden changes occur. The jumps take place at prime numbers to avoid an overly quick response due to matching the window size to the jump. In between, the scenario shows a parameter drift. Thus, the algorithm is evaluated for both, namely the adaption to drifts and jumps.
In comparison, Fig.~\ref{fig:SimIdentificationSL} shows the identification result of the proposed SL-based algorithm. It can be seen that the algorithm can track the statistical parameters with higher accuracy while it responds equally quick to the jumps. On top of that, the new algorithm features a measure of the statistical uncertainty which is depicted in Fig.~\ref{fig:SimIdentificationSL}(b). The peaks in the uncertainty result from the inconsistency of the opinions at the jumps which is detected by the consistency test.

\section{EXPERIMENTS} \label{sec:Experiment}
In our setup, cooperative information from a traffic monitoring system, which observes an occluded intersection, is sent to a Multi-Edge-Computing (MEC) server that computes an environment model from the received data and forwards the model to the CAV via a 5G test mobile network \cite{Buchholz2018}. To merge safely into the occluded intersection, the CAV needs a reliable, low-latency communication link to the MEC server. Thus, the reliability of the communication link must be estimated.

The idea of the presented approach is to infer the reliability of the communication link from the delay times the packets need to travel from the MEC server to the CAV. Therefore, the delays, i.e., the time differences between the timestamps when the MEC server sends a packet and the CAV receives the packet, are monitored.

The test mobile network uses also an HARQ method for error control. Thus, if a packet cannot be decoded immediately, a negative acknowledge (NACK) will be transmitted to request additional redundancy information to decode the packet at the second try \cite{Sauter2014}. This manifests in an additional delay of about $7 \, \text{ms}$. If the packet cannot be decoded the second time, further delays will occur.

To infer the reliability of the communication link, we first estimate whether a packet has been decoded immediately or had to be reconstructed through HARQ or further error correction mechanisms.
We model the communication channel as a three state Markov chain: \textit{state~1} denotes that a packet is received correctly, \textit{state~2} means that a packet needs HARQ correction which corresponds to slightly distorted conditions, and \textit{state~3} implies that a packet needs further correction mechanisms which corresponds to heavily distorted conditions. 

\begin{figure}[bt]
	\centering
	\includegraphics[width=0.5\textwidth]{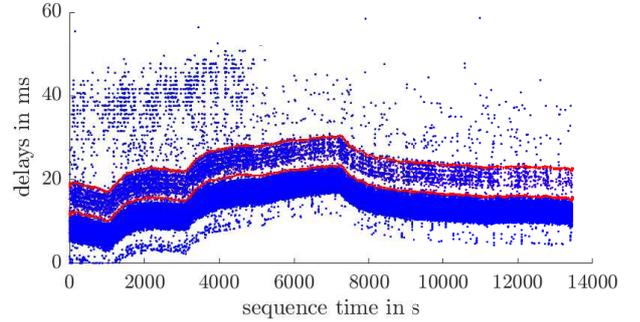}
	\caption{Recorded delays of a sequence of $127\,400$ packets. The decision thresholds are marked in red.}
	\label{fig:RecordedDelays}
\end{figure}

Note, however, that the delay of the received packets, even if they are successfully decoded at the first try, depends on multiple factors and thus can slowly vary over time. Therefore, the thresholds, with which the packets are classified as explained above, have to be adapted online. This is done by filtering the recently recorded delays with a moving average filter where outliers are excluded. To achieve a good decision threshold, some additional time is added to the moving average. Fig.~\ref{fig:RecordedDelays} shows a sequence of measured packet delays between the MEC server and the CAV as well as the decision borders. It can be seen that most of the packets are within the first region, i.e., below the lower red line, while some are in the second region, i.e., between the red lines, and only very few are in the third region, i.e., above the upper red line.

To evaluate these delays quantitatively, delays of about $127\,400$ packets were recorded over approximately $3$ hours. The sequence is subdivided into subsets of $l_w = 100$ recorded packet delays. The delays within these subsets are classified into \textit{immediately decoded}, \textit{HARQ corrected}, and \textit{further corrected}. The statistics over these subsets are monitored over time. These statistics represent the observed state transition rates of the Markov chain.

\begin{figure}[bthp]
	\centering
	\begin{tikzpicture}
	\footnotesize
	\input{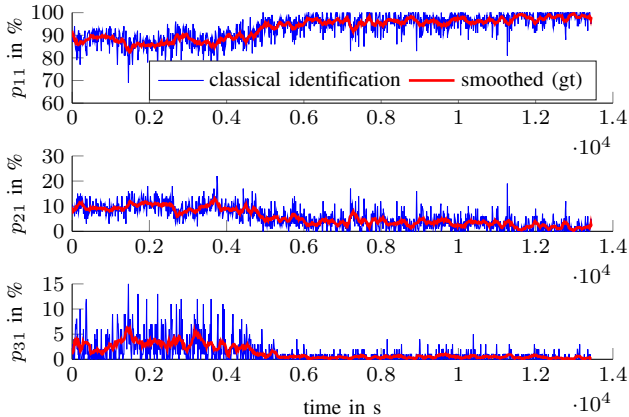}
	\end{tikzpicture}
	\caption{Classical statistical evaluation of the sequence over subsets with $l_w = 100$ samples and the smoothed result that serves as ground truth (gt).}
	\label{fig:GroundTruth}
\end{figure}

Fig.~\ref{fig:GroundTruth} shows the classically estimated state transition rates corresponding to the statistics as well as a smoothed version thereof. It can be seen that the probability of decoding a package at the first try is around $90 \, \%$ at the beginning of the sequence and increases afterwards to $95 \, \%$. This suits well with the typical radio link configuration. 
The increase of the decoding probability at the first try can be explained by a heavy rain shower that went over the test mobile network at the beginning of the sequence. The smoothed result of the classical identification is obtained with a moving average filter and is used later as ground truth.

\begin{figure}
	\subfloat[Identification result.]{
		\begin{tikzpicture}
		\footnotesize
		\input{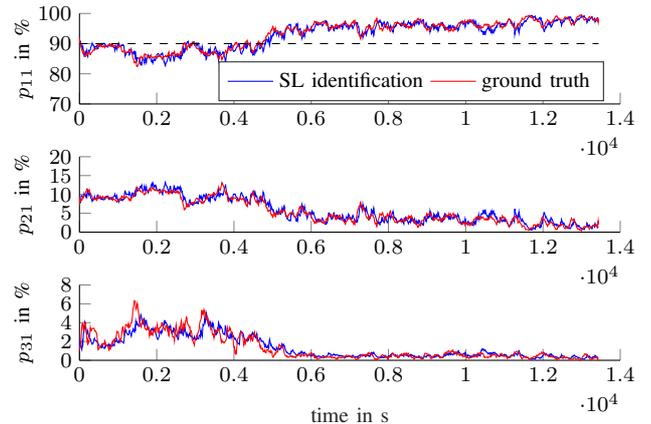}
		\end{tikzpicture}
	} \\
	\subfloat[Uncertainty.]{
		\begin{tikzpicture}
		\footnotesize
		\input{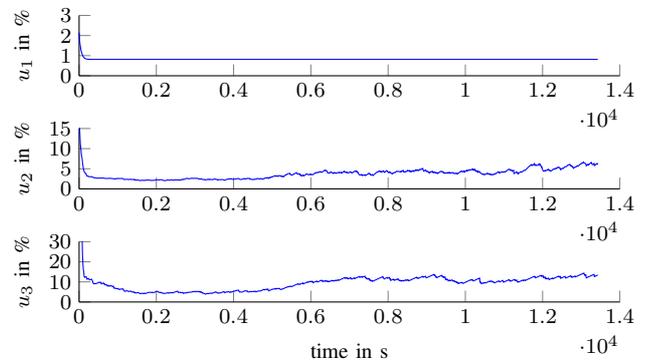}
		\end{tikzpicture}
	}
	\caption{This figure shows (a) the identification result of Algorithm \ref{alg:Algorithm} in comparison to the standard procedure and (b) the statistical uncertainty determined by the new SL-based algorithm. }
	\label{fig:Identification}
\end{figure}

In the next step, the Markov chain is identified using Algorithm \ref{alg:Algorithm} on the same data. The identification result $\boldsymbol{\Omega}$ is subsequently projected back to the state transition probabilities, as identified with the classical method. Both results are compared in Fig.~\ref{fig:Identification}(a), while the statistical uncertainty, which can only be obtained with the new method, is visualized in Fig.~\ref{fig:Identification}(b). It shows that the identification result of the proposed online method fits well with the acausally smoothed classical identification, and additionally delivers its uncertainty.

\section{CONCLUSIONS} \label{sec:Conclusion}
In this work, we proposed a SL-based method to identify Markov chains that integrates Markov chains into the theory of SL. The method has been applied to the reliability estimation of communication channels from CAVs. As reliable communication links are essential for the safety of CAVs, the proposed method is able to increase CAVs' safety. The method was evaluated through simulated data and validated by a real-world experiment showing that the method is able to track parameter changes in the communication channel.
In further work, the method will be tested on other network topologies. % such as G5.

%%%%%%%%%%%%%%%%%%%%%%%%%%%%%%%%%%%%%%%%%%%%%%%%%%%%%%%%%%%%%%%%%%%%%%%%%%%%%%%%

\section*{ACKNOWLEDGMENT}

We thank Soheil Gherekhloo for the fruitful discussions that helped us to review and improve our work.

%%%%%%%%%%%%%%%%%%%%%%%%%%%%%%%%%%%%%%%%%%%%%%%%%%%%%%%%%%%%%%%%%%%%%%%%%%%%%%%%

\bibliographystyle{IEEEtran}
{
	\bibliography{CAVS2019}}

\end{document}

%% file: img/ClassicalIdentificationSim.tex
% This file was created by matlab2tikz.
%
%The latest updates can be retrieved from
%  http://www.mathworks.com/matlabcentral/fileexchange/22022-matlab2tikz-matlab2tikz
%where you can also make suggestions and rate matlab2tikz.
%
%\begin{tikzpicture}

\begin{axis}[%
width=0.4 \textwidth,
height=1.5cm,
at={(0cm,1.7cm)},
scale only axis,
xmin=0,
xmax=100000,
ymin=40,
ymax=100,
y label style={at={(axis description cs:0.07,0.5)}},
ylabel={$p_{GG}$ in \%},
axis background/.style={fill=white},
axis x line*=bottom,
axis y line*=left,
legend style={at={(0.97,0.03)}, anchor=south east, legend columns=2, legend cell align=left, align=left, draw=white!15!black}
]
\addplot [color=blue]
  table[row sep=crcr]{%
100	80.246913580253\\
200	86.9047619047633\\
300	96.875\\
400	93.4065934065875\\
500	91.0112359550549\\
600	87.3563218390773\\
700	89.8876404494367\\
800	88.7640449438186\\
900	85.3658536585426\\
1000	93.4782608695677\\
1100	93.3333333333285\\
1200	88.5057471264299\\
1300	89.7727272727207\\
1400	88.7640449438186\\
1500	93.3333333333285\\
1600	88.6363636363676\\
1700	93.4065934065875\\
1800	94.6808510638366\\
1900	89.7727272727207\\
2000	82.5\\
2100	92.3076923076878\\
2200	88.0952380952367\\
2300	93.4782608695677\\
2400	88.2352941176505\\
2500	85.1851851851825\\
2600	88.2352941176505\\
2700	87.2093023255875\\
2800	93.1818181818235\\
2900	92.222222222219\\
3000	88.7640449438186\\
3100	85.8823529411748\\
3200	91.0112359550549\\
3300	93.4065934065875\\
3400	84.5238095238165\\
3500	93.5483870967728\\
3600	89.8876404494367\\
3700	89.6551724137971\\
3800	88.7640449438186\\
3900	88.6363636363676\\
4000	89.7727272727207\\
4100	92.3913043478242\\
4200	90.5882352941117\\
4300	89.8876404494367\\
4400	96.875\\
4500	92.0454545454559\\
4600	87.3563218390773\\
4700	87.3563218390773\\
4800	93.4782608695677\\
4900	91.2087912087882\\
5000	90.9090909090883\\
5100	91.1111111111095\\
5200	91.0112359550549\\
5300	88.2352941176505\\
5400	94.5054945055017\\
5500	93.5483870967728\\
5600	92.134831460673\\
5700	88.3720930232521\\
5800	92.134831460673\\
5900	88.2352941176505\\
6000	89.7727272727207\\
6100	89.6551724137971\\
6200	96.8421052631602\\
6300	92.134831460673\\
6400	88.0952380952367\\
6500	87.3563218390773\\
6600	92.3913043478242\\
6700	87.3563218390773\\
6800	95.6043956044014\\
6900	95.6043956044014\\
7000	85.7142857142899\\
7100	94.6236559139797\\
7200	89.8876404494367\\
7300	92.3076923076878\\
7400	95.6521739130403\\
7500	90.1098901098885\\
7600	87.5\\
7700	92.3076923076878\\
7800	93.5483870967728\\
7900	88.5057471264299\\
8000	87.2093023255875\\
8100	87.2093023255875\\
8200	94.6236559139797\\
8300	95.6989247311867\\
8400	87.8048780487734\\
8500	91.0112359550549\\
8600	92.0454545454559\\
8700	85.7142857142899\\
8800	87.2093023255875\\
8900	88.5057471264299\\
9000	92.222222222219\\
9100	88.5057471264299\\
9200	92.222222222219\\
9300	88.3720930232521\\
9400	94.5054945055017\\
9500	92.222222222219\\
9600	86.0759493670921\\
9700	86.0759493670921\\
9800	86.9047619047633\\
9900	90.9090909090883\\
10000	85.3658536585426\\
10100	94.5054945055017\\
10200	88.0952380952367\\
10300	89.7727272727207\\
10400	90.8045977011498\\
10500	92.134831460673\\
10600	91.2087912087882\\
10700	90.5882352941117\\
10800	93.4065934065875\\
10900	92.134831460673\\
11000	94.6808510638366\\
11100	89.7727272727207\\
11200	91.2087912087882\\
11300	89.6551724137971\\
11400	92.3076923076878\\
11500	83.75\\
11600	87.2093023255875\\
11700	92.3076923076878\\
11800	92.3076923076878\\
11900	86.746987951803\\
12000	89.0243902438961\\
12100	89.5348837209312\\
12200	89.6551724137971\\
12300	92.134831460673\\
12400	86.0465116279083\\
12500	91.1111111111095\\
12600	82.5\\
12700	89.7727272727207\\
12800	85.1851851851825\\
12900	87.0588235294126\\
13000	90.9090909090883\\
13100	94.6808510638366\\
13200	93.4065934065875\\
13300	78.4810126582306\\
13400	90.9090909090883\\
13500	93.4782608695677\\
13600	94.6808510638366\\
13700	89.8876404494367\\
13800	90\\
13900	92.3076923076878\\
14000	89.5348837209312\\
14100	91.1111111111095\\
14200	85.8823529411748\\
14300	92.134831460673\\
14400	86.9047619047633\\
14500	93.5483870967728\\
14600	89.4117647058883\\
14700	92.3076923076878\\
14800	88.6363636363676\\
14900	88.6363636363676\\
15000	88.7640449438186\\
15100	88.3720930232521\\
15200	95.7894736842136\\
15300	87.9518072289211\\
15400	90.5882352941117\\
15500	89.8876404494367\\
15600	92.3076923076878\\
15700	88.7640449438186\\
15800	92.134831460673\\
15900	88.6363636363676\\
16000	92.3913043478242\\
16100	88.0952380952367\\
16200	89.4117647058883\\
16300	94.3820224719093\\
16400	88.2352941176505\\
16500	89.5348837209312\\
16600	82.6666666666715\\
16700	88.6363636363676\\
16800	89.7727272727207\\
16900	91.0112359550549\\
17000	90.9090909090883\\
17100	84.6153846153902\\
17200	96.8421052631602\\
17300	90.1098901098885\\
17400	92.3076923076878\\
17500	88.3720930232521\\
17600	85.5263157894806\\
17700	89.1566265060246\\
17800	92.3913043478242\\
17900	88.7640449438186\\
18000	91.860465116275\\
18100	88.3720930232521\\
18200	87.3563218390773\\
18300	92.3076923076878\\
18400	93.617021276601\\
18500	82.5\\
18600	88.3720930232521\\
18700	90.4761904761835\\
18800	83.9506172839465\\
18900	89.6551724137971\\
19000	85.3658536585426\\
19100	63.076923076922\\
19200	76\\
19300	79.4871794871724\\
19400	76.6233766233781\\
19500	70.8333333333285\\
19600	69.444444444438\\
19700	71.4285714285652\\
19800	76.6233766233781\\
19900	70.4225352112699\\
20000	66.1764705882379\\
20100	74.6666666666715\\
20200	67.6470588235243\\
20300	73.3333333333285\\
20400	75\\
20500	72.8571428571449\\
20600	67.6923076923122\\
20700	80.5194805194769\\
20800	73.6111111111095\\
20900	69.8630136986321\\
21000	66.6666666666715\\
21100	75.3424657534197\\
21200	62.5\\
21300	70\\
21400	74.6666666666715\\
21500	71.8309859154979\\
21600	77.6315789473738\\
21700	73.2394366197113\\
21800	67.6470588235243\\
21900	71.8309859154979\\
22000	73.2394366197113\\
22100	70.7692307692341\\
22200	76.923076923078\\
22300	60.3174603174557\\
22400	71.6216216216271\\
22500	74.3243243243196\\
22600	69.0140845070418\\
22700	77.2151898734155\\
22800	68.5714285714348\\
22900	71.6216216216271\\
23000	75.3246753246785\\
23100	54.2372881355986\\
23200	61.6666666666715\\
23300	81.9277108433744\\
23400	61.9047619047633\\
23500	71.4285714285652\\
23600	69.8630136986321\\
23700	73.3333333333285\\
23800	71.4285714285652\\
23900	76.0563380281674\\
24000	64.1791044776182\\
24100	70.1492537313461\\
24200	72.6027397260332\\
24300	81.4814814814745\\
24400	80\\
24500	65.6716417910502\\
24600	65.0793650793639\\
24700	70.4225352112699\\
24800	67.1641791044822\\
24900	69.444444444438\\
25000	74.2857142857101\\
25100	68.6567164179141\\
25200	72.4637681159365\\
25300	73.9726027397264\\
25400	70.8333333333285\\
25500	62.1212121212156\\
25600	78.6666666666715\\
25700	73.6111111111095\\
25800	66.6666666666715\\
25900	60.6557377049176\\
26000	70.4225352112699\\
26100	75.3623188405763\\
26200	64.6153846153902\\
26300	75.3424657534197\\
26400	78.2051282051252\\
26500	66.1538461538439\\
26600	71.2328767123254\\
26700	73.3333333333285\\
26800	72.222222222219\\
26900	66.1538461538439\\
27000	73.9726027397264\\
27100	63.6363636363676\\
27200	75.3424657534197\\
27300	72.222222222219\\
27400	67.7419354838639\\
27500	78.0821917808207\\
27600	71.4285714285652\\
27700	73.2394366197113\\
27800	71.2328767123254\\
27900	68.6567164179141\\
28000	74.626865671642\\
28100	71.6417910447781\\
28200	82.9268292682973\\
28300	67.6470588235243\\
28400	73.9130434782564\\
28500	58.4615384615317\\
28600	68.6567164179141\\
28700	74.6666666666715\\
28800	76.3888888888905\\
28900	79.7468354430312\\
29000	79.4871794871724\\
29100	75.7142857142899\\
29200	76.7123287671275\\
29300	78.4810126582306\\
29400	70.5882352941117\\
29500	82.9268292682973\\
29600	67.6470588235243\\
29700	77.4647887323954\\
29800	70.8333333333285\\
29900	71.4285714285652\\
30000	64.1791044776182\\
30100	78.4810126582306\\
30200	69.5652173913113\\
30300	75.3424657534197\\
30400	72.222222222219\\
30500	78.3783783783729\\
30600	60.9375\\
30700	68.055555555562\\
30800	68.852459016387\\
30900	68.6274509803916\\
31000	75.3846153846098\\
31100	73.3333333333285\\
31200	71.4285714285652\\
31300	65.3061224489793\\
31400	68.0851063829759\\
31500	64.7058823529369\\
31600	78.6885245901649\\
31700	74.626865671642\\
31800	70.5882352941117\\
31900	77.777777777781\\
32000	66.0377358490514\\
32100	66.6666666666715\\
32200	79.6875\\
32300	68.852459016387\\
32400	84.5070422535209\\
32500	77.1929824561375\\
32600	76.2711864406738\\
32700	72.7272727272793\\
32800	70.5882352941117\\
32900	66.6666666666715\\
33000	69.6428571428551\\
33100	82.6086956521758\\
33200	77.9411764705874\\
33300	66.0714285714348\\
33400	67.8571428571449\\
33500	84.2105263157864\\
33600	79.3650793650741\\
33700	69.4915254237276\\
33800	71.9298245614045\\
33900	69.0909090909117\\
34000	77.1929824561375\\
34100	71.4285714285652\\
34200	82.2580645161361\\
34300	76.2711864406738\\
34400	72.4137931034493\\
34500	69.4915254237276\\
34600	66.1016949152545\\
34700	71.1538461538439\\
34800	70.4918032786954\\
34900	79.1044776119379\\
35000	63.4615384615317\\
35100	75\\
35200	71.9298245614045\\
35300	72.7272727272793\\
35400	71.6666666666715\\
35500	69.0909090909117\\
35600	62.5\\
35700	67.2131147540931\\
35800	76.923076923078\\
35900	78.125\\
36000	73.6842105263204\\
36100	77.2727272727207\\
36200	71.1538461538439\\
36300	66.6666666666715\\
36400	62.222222222219\\
36500	60.7843137254968\\
36600	72.4137931034493\\
36700	72.7272727272793\\
36800	68.2539682539646\\
36900	69.6428571428551\\
37000	75\\
37100	64.1509433962201\\
37200	72\\
37300	72.580645161288\\
37400	79.1044776119379\\
37500	76.1904761904734\\
37600	74.1935483870911\\
37700	65.5737704917992\\
37800	71.6666666666715\\
37900	78.5714285714348\\
38000	74.626865671642\\
38100	70.1754385964887\\
38200	87.1428571428551\\
38300	74.626865671642\\
38400	77.3584905660391\\
38500	74.5762711864372\\
38600	73.4375\\
38700	71.4285714285652\\
38800	66.1538461538439\\
38900	75.3846153846098\\
39000	71.6981132075452\\
39100	78.3783783783729\\
39200	77.2727272727207\\
39300	74.6031746031804\\
39400	79.7297297297337\\
39500	81.9672131147527\\
39600	82.3529411764757\\
39700	72.4137931034493\\
39800	77.6119402985059\\
39900	73.7704918032832\\
40000	68.5185185185255\\
40100	66.6666666666715\\
40200	76.1904761904734\\
40300	85.3333333333285\\
40400	87.1794871794846\\
40500	75\\
40600	73.7704918032832\\
40700	75.8064516129089\\
40800	67.2413793103478\\
40900	75.4098360655771\\
41000	78.3333333333285\\
41100	65.4545454545441\\
41200	82.0895522388018\\
41300	76.6666666666715\\
41400	81.5384615384683\\
41500	82.3529411764757\\
41600	69.3877551020414\\
41700	68.6274509803916\\
41800	77.6119402985059\\
41900	77.5862068965507\\
42000	71.9298245614045\\
42100	68.5185185185255\\
42200	73.3333333333285\\
42300	65.9574468085048\\
42400	84\\
42500	77.419354838712\\
42600	76.470588235301\\
42700	82.6086956521758\\
42800	74.5762711864372\\
42900	61.2244897959172\\
43000	73.3333333333285\\
43100	76.5625\\
43200	75.7575757575687\\
43300	70.1754385964887\\
43400	73.4375\\
43500	77.419354838712\\
43600	71.9298245614045\\
43700	60\\
43800	72.4137931034493\\
43900	74.1935483870911\\
44000	83.076923076922\\
44100	76.5625\\
44200	77.777777777781\\
44300	77.9411764705874\\
44400	72.8813559322007\\
44500	78.7878787878726\\
44600	75.4716981132078\\
44700	74.6031746031804\\
44800	72.8813559322007\\
44900	66.6666666666715\\
45000	81.5384615384683\\
45100	78.4615384615317\\
45200	81.6901408450649\\
45300	84.6153846153902\\
45400	84\\
45500	87.3417721518927\\
45600	76.470588235301\\
45700	75.4098360655771\\
45800	76.2711864406738\\
45900	75.4098360655771\\
46000	70.2127659574471\\
46100	71.875\\
46200	88.1578947368398\\
46300	81.4285714285652\\
46400	73.3333333333285\\
46500	79.710144927536\\
46600	82.2580645161361\\
46700	75.3846153846098\\
46800	74.1379310344782\\
46900	87.323943661977\\
47000	87.8378378378402\\
47100	76.5625\\
47200	76.6666666666715\\
47300	80.3278688524588\\
47400	75.3846153846098\\
47500	75.3623188405763\\
47600	76.5625\\
47700	85.2941176470631\\
47800	73.2142857142899\\
47900	76.5625\\
48000	75.7575757575687\\
48100	68.5185185185255\\
48200	73.5849056603765\\
48300	62.7450980392168\\
48400	83.3333333333285\\
48500	70.9677419354848\\
48600	71.6666666666715\\
48700	85.5263157894806\\
48800	76.119402985074\\
48900	77.4647887323954\\
49000	83.5820895522338\\
49100	81.5384615384683\\
49200	77.6119402985059\\
49300	76.923076923078\\
49400	71.875\\
49500	83.3333333333285\\
49600	74.6031746031804\\
49700	79.3650793650741\\
49800	77.9661016949103\\
49900	84.8484848484804\\
50000	75\\
50100	81.6901408450649\\
50200	67.2413793103478\\
50300	83.3333333333285\\
50400	85.9154929577489\\
50500	71.4285714285652\\
50600	79.3650793650741\\
50700	73.9130434782564\\
50800	78.5714285714348\\
50900	76.1904761904734\\
51000	72.8813559322007\\
51100	78.2608695652161\\
51200	75.4098360655771\\
51300	85.1351351351332\\
51400	73.2142857142899\\
51500	81.4285714285652\\
51600	79.6875\\
51700	83.8235294117621\\
51800	71.4285714285652\\
51900	72.222222222219\\
52000	76.923076923078\\
52100	76.470588235301\\
52200	81.1594202898559\\
52300	72.4137931034493\\
52400	81.0810810810799\\
52500	80.5970149253699\\
52600	70.6896551724203\\
52700	75\\
52800	82.191780821915\\
52900	75\\
53000	71.4285714285652\\
53100	73.8461538461561\\
53200	75.4385964912217\\
53300	70.1754385964887\\
53400	81.6901408450649\\
53500	78.9473684210534\\
53600	79.710144927536\\
53700	70.1754385964887\\
53800	86.4864864864794\\
53900	73.5849056603765\\
54000	79.3650793650741\\
54100	81.1594202898559\\
54200	72.580645161288\\
54300	79.3650793650741\\
54400	75.7575757575687\\
54500	72.222222222219\\
54600	73.2142857142899\\
54700	75.8064516129089\\
54800	76.1904761904734\\
54900	80.555555555562\\
55000	76.2711864406738\\
55100	72.4137931034493\\
55200	80.3030303030246\\
55300	72.1311475409893\\
55400	70.370370370365\\
55500	80.9523809523816\\
55600	88\\
55700	83.3333333333285\\
55800	68.6274509803916\\
55900	83.5616438356228\\
56000	80.3278688524588\\
56100	76.119402985074\\
56200	79.0322580645152\\
56300	82.8947368421068\\
56400	77.419354838712\\
56500	81.8181818181765\\
56600	83.5820895522338\\
56700	75.3846153846098\\
56800	71.6666666666715\\
56900	84.722222222219\\
57000	82.6666666666715\\
57100	75.5102040816273\\
57200	75\\
57300	72.222222222219\\
57400	84.5070422535209\\
57500	72.4137931034493\\
57600	71.4285714285652\\
57700	80\\
57800	79.4117647058883\\
57900	71.1864406779641\\
58000	76.1904761904734\\
58100	76.3888888888905\\
58200	80.5970149253699\\
58300	80\\
58400	75.7575757575687\\
58500	74.5762711864372\\
58600	80.8823529411748\\
58700	75.8620689655218\\
58800	80.7017543859693\\
58900	77.9411764705874\\
59000	80.5970149253699\\
59100	87.8048780487734\\
59200	80.2816901408514\\
59300	76.8115942028962\\
59400	81.5384615384683\\
59500	83.1168831168761\\
59600	85.9154929577489\\
59700	79.4117647058883\\
59800	75.7142857142899\\
59900	84.810126582277\\
60000	81.4285714285652\\
60100	84.7457627118711\\
60200	83.0985915492929\\
60300	81.8181818181765\\
60400	78.8732394366234\\
60500	77.9411764705874\\
60600	82.3529411764757\\
60700	90.2439024390187\\
60800	81.1594202898559\\
60900	77.9411764705874\\
61000	82.4324324324261\\
61100	77.2727272727207\\
61200	78.5714285714348\\
61300	80.2816901408514\\
61400	79.1044776119379\\
61600	79.710144927536\\
61700	84\\
61800	84.5070422535209\\
61900	80.5970149253699\\
62000	80.3030303030246\\
62100	75\\
62200	76.119402985074\\
62300	83.3333333333285\\
62400	82.8947368421068\\
62500	87.5\\
62600	81.6901408450649\\
62700	80\\
62800	77.2727272727207\\
62900	76.470588235301\\
63000	85.7142857142899\\
63100	84.0579710144957\\
63200	75\\
63300	82.8947368421068\\
63400	75.3846153846098\\
63500	77.777777777781\\
63600	84.4155844155903\\
63700	75.8064516129089\\
63800	76.5625\\
63900	76.119402985074\\
64000	84.5070422535209\\
64100	81.6901408450649\\
64200	80\\
64300	89.0243902438961\\
64400	84.6153846153902\\
64500	81.5789473684272\\
64600	86.3013698630093\\
64700	83.3333333333285\\
64800	74.626865671642\\
64900	87.0129870129895\\
65000	80.5194805194769\\
65100	80.9523809523816\\
65200	83.5443037974619\\
65300	84.4155844155903\\
65400	80.7692307692341\\
65500	58.3333333333285\\
65600	79.1666666666715\\
65700	81.8181818181765\\
65800	87.8378378378402\\
65900	84.2105263157864\\
66000	84\\
66100	76.3888888888905\\
66200	90.3614457831281\\
66300	84.4155844155903\\
66400	75.7142857142899\\
66500	82.3529411764757\\
66600	75.3623188405763\\
66700	71.8309859154979\\
66800	86.5853658536653\\
66900	82.4324324324261\\
67000	84.722222222219\\
67100	86.9047619047633\\
67200	85.9154929577489\\
67300	91.860465116275\\
67400	86.4197530864185\\
67500	82.8125\\
67600	79.1666666666715\\
67700	92.222222222219\\
67800	84.4155844155903\\
67900	81.5789473684272\\
68000	85.1851851851825\\
68100	88.4615384615317\\
68200	84.2105263157864\\
68300	77.1428571428551\\
68400	83.5820895522338\\
68500	82.6666666666715\\
68600	83.75\\
68700	81.1594202898559\\
68800	88.4615384615317\\
68900	83.7837837837869\\
69000	76.470588235301\\
69100	71.4285714285652\\
69200	84.4155844155903\\
69300	86.9047619047633\\
69400	73.9130434782564\\
69500	75\\
69600	80.8219178082218\\
69700	76.3888888888905\\
69800	91.860465116275\\
69900	80\\
70000	85.1851851851825\\
70100	79.1666666666715\\
70200	75.3623188405763\\
70300	74.6478873239394\\
70400	87.1794871794846\\
70500	85.7142857142899\\
70600	81.5789473684272\\
70700	81.4285714285652\\
70800	87.9518072289211\\
70900	84.722222222219\\
71000	81.3333333333285\\
71100	85.5263157894806\\
71200	81.1594202898559\\
71300	77.777777777781\\
71400	78.3783783783729\\
71500	81.3333333333285\\
71600	83.5820895522338\\
71700	90\\
71800	76.0563380281674\\
71900	89.8734177215229\\
72000	80\\
72100	89.2857142857101\\
72200	83.5443037974619\\
72300	81.5789473684272\\
72400	86.5853658536653\\
72500	79.1666666666715\\
72600	78.9473684210534\\
72700	80.2816901408514\\
72800	86.9047619047633\\
72900	82.5396825396892\\
73000	87.3417721518927\\
73100	80.555555555562\\
73200	80.8219178082218\\
73300	84.4155844155903\\
73400	80.7692307692341\\
73500	79.7297297297337\\
73600	90.2439024390187\\
73700	75.7142857142899\\
73800	80.5194805194769\\
73900	87.6543209876545\\
74000	84.14634146342\\
74100	86.0759493670921\\
74200	83.5443037974619\\
74300	81.8181818181765\\
74400	85.5263157894806\\
74500	82.6086956521758\\
74600	78.2608695652161\\
74700	70.7692307692341\\
74800	82.0512820512813\\
74900	77.0270270270266\\
75000	83.3333333333285\\
75100	84.6153846153902\\
75200	86.5853658536653\\
75300	76.3888888888905\\
75400	82.6666666666715\\
75500	77.6315789473738\\
75600	85.5263157894806\\
75700	79.7297297297337\\
75800	78.5714285714348\\
75900	86.9047619047633\\
76000	83.75\\
76100	81.3333333333285\\
76200	81.1594202898559\\
76300	87.3563218390773\\
76400	86.4197530864185\\
76500	81.944444444438\\
76600	73.9726027397264\\
76700	76.3888888888905\\
76800	82.5\\
76900	82.8947368421068\\
77000	90.4761904761835\\
77100	82.8947368421068\\
77200	79.7297297297337\\
77300	85.1351351351332\\
77400	84.4155844155903\\
77500	81.8181818181765\\
77600	84.14634146342\\
77700	85.3658536585426\\
77800	78.8732394366234\\
77900	85.8823529411748\\
78000	77.1428571428551\\
78100	85.3658536585426\\
78200	86.25\\
78300	75\\
78400	80.263157894733\\
78500	83.5443037974619\\
78600	86.5853658536653\\
78700	82.0512820512813\\
78800	77.1428571428551\\
78900	82.9268292682973\\
79000	80.263157894733\\
79100	83.3333333333285\\
79200	74.2857142857101\\
79300	75\\
79400	89.2857142857101\\
79500	87.0588235294126\\
79600	73.9726027397264\\
79700	84.337349397596\\
79800	83.7837837837869\\
79900	91.5662650602462\\
80000	87.0588235294126\\
80100	90.8045977011498\\
80200	82.0512820512813\\
80300	81.5789473684272\\
80400	86.9047619047633\\
80500	88.6075949367078\\
80600	85.5421686746995\\
80700	89.0243902438961\\
80800	89.5348837209312\\
80900	85.7142857142899\\
81000	89.7727272727207\\
81100	87.0588235294126\\
81200	86.0759493670921\\
81300	89.4117647058883\\
81400	83.3333333333285\\
81500	83.75\\
81600	91.860465116275\\
81700	82.5\\
81800	87.8048780487734\\
81900	83.3333333333285\\
82000	86.25\\
82100	79.1044776119379\\
82200	91.0112359550549\\
82300	89.0243902438961\\
82400	84.810126582277\\
82500	90.1234567901265\\
82600	85.8823529411748\\
82700	84.14634146342\\
82800	85.8823529411748\\
82900	83.3333333333285\\
83000	85\\
83100	85.8974358974374\\
83200	93.2584269662912\\
83300	85.1851851851825\\
83400	86.746987951803\\
83500	89.2857142857101\\
83600	84.5238095238165\\
83700	86.9047619047633\\
83800	89.6551724137971\\
83900	89.2857142857101\\
84000	86.5853658536653\\
84100	91.5662650602462\\
84200	89.5348837209312\\
84300	83.1168831168761\\
84400	78.6666666666715\\
84500	87.8048780487734\\
84600	88.6363636363676\\
84700	85.3658536585426\\
84800	91.2087912087882\\
84900	90.6976744186104\\
85000	87.0588235294126\\
85100	90.9090909090883\\
85200	87.0588235294126\\
85300	83.9506172839465\\
85400	87.6543209876545\\
85500	89.4117647058883\\
85600	87.0588235294126\\
85700	82.8947368421068\\
85800	80.555555555562\\
85900	87.3417721518927\\
86000	83.9506172839465\\
86100	85\\
86200	89.5348837209312\\
86300	83.3333333333285\\
86400	90.5882352941117\\
86500	79.4871794871724\\
86600	80.263157894733\\
86700	82.0512820512813\\
86800	84.810126582277\\
86900	92.134831460673\\
87000	87.9518072289211\\
87100	88.75\\
87200	81.3333333333285\\
87300	90.8045977011498\\
87400	81.25\\
87500	93.2584269662912\\
87600	89.4117647058883\\
87700	87.9518072289211\\
87800	86.746987951803\\
87900	89.4117647058883\\
88000	80.246913580253\\
88100	88.2352941176505\\
88200	85.7142857142899\\
88300	87.8048780487734\\
88400	86.9047619047633\\
88500	85.1851851851825\\
88600	82.5\\
88700	87.0588235294126\\
88800	93.2584269662912\\
88900	89.5348837209312\\
89000	83.1168831168761\\
89100	91.0112359550549\\
89200	79.4871794871724\\
89300	85.8823529411748\\
89400	85\\
89500	87.2093023255875\\
89600	81.0126582278463\\
89700	92.3076923076878\\
89800	81.4814814814745\\
89900	89.0243902438961\\
90000	84.7058823529369\\
90100	88.6075949367078\\
90200	86.746987951803\\
90300	90.3614457831281\\
90400	81.8181818181765\\
90500	84.14634146342\\
90600	81.4814814814745\\
90700	84.6153846153902\\
90800	91.860465116275\\
90900	88.2352941176505\\
91000	89.8876404494367\\
91100	92.222222222219\\
91200	89.7727272727207\\
91300	86.746987951803\\
91400	84.810126582277\\
91500	94.6808510638366\\
91600	91.9540229885024\\
91700	87.6543209876545\\
91800	77.2151898734155\\
91900	84.14634146342\\
92000	82.0512820512813\\
92100	84.14634146342\\
92200	90.8045977011498\\
92300	89.5348837209312\\
92400	88.6363636363676\\
92500	92.0454545454559\\
92600	82.5\\
92700	84.14634146342\\
92800	84.5238095238165\\
92900	93.5483870967728\\
93000	82.9268292682973\\
93100	84.810126582277\\
93200	85.8823529411748\\
93300	81.4814814814745\\
93400	85.7142857142899\\
93500	82.2784810126614\\
93600	85.8823529411748\\
93700	93.4782608695677\\
93800	77.9220779220777\\
93900	87.2093023255875\\
94000	86.9047619047633\\
94100	86.9047619047633\\
94200	80.246913580253\\
94300	81.8181818181765\\
94400	94.444444444438\\
94500	90.6976744186104\\
94600	82.5\\
94700	86.746987951803\\
94800	90.8045977011498\\
94900	83.1325301204779\\
95000	85.8974358974374\\
95100	90.9090909090883\\
95200	95.4022988505749\\
95300	83.9506172839465\\
95400	88.6363636363676\\
95500	89.4117647058883\\
95600	85.5421686746995\\
95700	83.1325301204779\\
95800	88.2352941176505\\
95900	81.4814814814745\\
96000	96.8421052631602\\
96100	93.5483870967728\\
96200	92.134831460673\\
96300	84.7058823529369\\
96400	92.3913043478242\\
96500	88.5057471264299\\
96600	87.0588235294126\\
96700	81.25\\
96800	89.5348837209312\\
96900	91.2087912087882\\
97000	84.337349397596\\
97100	89.6551724137971\\
97200	89.4117647058883\\
97300	91.2087912087882\\
97400	82.9268292682973\\
97500	85.8823529411748\\
97600	76.3157894736796\\
97700	91.860465116275\\
97800	88.5057471264299\\
97900	91.3043478260806\\
98000	85\\
98100	85.1851851851825\\
98200	88.5057471264299\\
98300	85.7142857142899\\
98400	91.0112359550549\\
98500	89.5348837209312\\
98600	97.8723404255288\\
98700	93.5483870967728\\
98800	92.3913043478242\\
98900	87.2093023255875\\
99000	89.8876404494367\\
99100	87.9518072289211\\
99200	92.134831460673\\
99300	86.0465116279083\\
99400	92.134831460673\\
99500	87.0588235294126\\
99600	91.9540229885024\\
99700	89.5348837209312\\
99800	91.1111111111095\\
99900	90\\
};
\addlegendentry{classical identification}

\addplot [color=red, line width=1.0pt]
  table[row sep=crcr]{%
100	90\\
19100	90\\
19200	70.0046961121116\\
100000	89.975530784257\\
};
\addlegendentry{ground truth}
\end{axis}

\begin{axis}[%
width=0.4 \textwidth,
height=1.0cm,
at={(0cm,0.0cm)},
scale only axis,
xmin=0,
xmax=100000,
x label style={at={(axis description cs:0.5,0.1)}},
xlabel={packets},
ymin=0,
ymax=80,
y label style={at={(axis description cs:0.07,0.5)}},
ylabel={$p_{BB}$ in \%},
axis background/.style={fill=white},
axis x line*=bottom,
axis y line*=left
]
\addplot [color=blue, forget plot]
  table[row sep=crcr]{%
100	11.1111111111095\\
200	26.6666666666715\\
300	0\\
400	25\\
500	30\\
600	8.33333333332848\\
700	10\\
800	0\\
900	29.4117647058883\\
1000	28.5714285714348\\
1100	33.3333333333285\\
1200	16.6666666666715\\
1300	27.2727272727207\\
1400	10\\
1500	33.3333333333285\\
1600	18.1818181818235\\
1700	25\\
1800	0\\
1900	18.1818181818235\\
2000	26.3157894736796\\
2100	12.5\\
2200	33.3333333333285\\
2300	14.2857142857101\\
2400	28.5714285714348\\
2500	27.777777777781\\
2600	35.7142857142899\\
2700	15.3846153846098\\
2800	36.3636363636324\\
2900	22.222222222219\\
3000	10\\
3100	14.2857142857101\\
3200	20\\
3300	25\\
3400	13.3333333333285\\
3500	0\\
3600	0\\
3700	25\\
3800	0\\
4000	18.1818181818235\\
4100	0\\
4200	35.7142857142899\\
4300	10\\
4400	0\\
4500	36.3636363636324\\
4600	8.33333333332848\\
4700	8.33333333332848\\
4800	14.2857142857101\\
4900	0\\
5000	18.1818181818235\\
5100	11.1111111111095\\
5200	20\\
5300	28.5714285714348\\
5400	25\\
5500	16.6666666666715\\
5600	30\\
5700	23.076923076922\\
5800	30\\
5900	28.5714285714348\\
6000	18.1818181818235\\
6100	25\\
6200	25\\
6300	30\\
6400	33.3333333333285\\
6500	8.33333333332848\\
6600	0\\
6700	8.33333333332848\\
6800	50\\
6900	50\\
7000	20\\
7100	0\\
7200	10\\
7300	12.5\\
7400	42.8571428571449\\
7500	0\\
7600	0\\
7700	12.5\\
7800	0\\
7900	16.6666666666715\\
8000	7.69230769231217\\
8100	15.3846153846098\\
8200	16.6666666666715\\
8300	33.3333333333285\\
8400	47.0588235294126\\
8500	20\\
8600	36.3636363636324\\
8700	20\\
8800	15.3846153846098\\
8900	16.6666666666715\\
9000	11.1111111111095\\
9200	22.222222222219\\
9300	23.076923076922\\
9400	37.5\\
9500	22.222222222219\\
9600	45\\
9700	45\\
9800	26.6666666666715\\
9900	27.2727272727207\\
10000	29.4117647058883\\
10100	37.5\\
10200	33.3333333333285\\
10300	18.1818181818235\\
10400	33.3333333333285\\
10500	30\\
10600	0\\
10700	42.8571428571449\\
10800	25\\
10900	30\\
11000	0\\
11100	18.1818181818235\\
11200	0\\
11300	25\\
11400	25\\
11500	31.5789473684272\\
11600	15.3846153846098\\
11700	12.5\\
11800	12.5\\
11900	31.25\\
12000	47.0588235294126\\
12100	30.7692307692341\\
12200	25\\
12300	30\\
12400	7.69230769231217\\
12500	11.1111111111095\\
12600	26.3157894736796\\
12700	18.1818181818235\\
12800	27.777777777781\\
12900	21.4285714285652\\
13000	27.2727272727207\\
13100	0\\
13200	25\\
13300	15\\
13400	27.2727272727207\\
13500	14.2857142857101\\
13600	0\\
13700	10\\
13800	0\\
13900	12.5\\
14000	30.7692307692341\\
14100	11.1111111111095\\
14200	14.2857142857101\\
14300	30\\
14400	26.6666666666715\\
14500	0\\
14600	28.5714285714348\\
14700	12.5\\
14800	9.09090909091174\\
14900	9.09090909091174\\
15000	0\\
15100	23.076923076922\\
15200	0\\
15300	37.5\\
15400	35.7142857142899\\
15500	10\\
15600	12.5\\
15700	0\\
15800	30\\
15900	18.1818181818235\\
16000	0\\
16100	33.3333333333285\\
16200	35.7142857142899\\
16300	40\\
16400	35.7142857142899\\
16500	30.7692307692341\\
16600	45.8333333333285\\
16700	9.09090909091174\\
16800	18.1818181818235\\
16900	20\\
17000	27.2727272727207\\
17100	38.0952380952367\\
17200	0\\
17300	0\\
17400	12.5\\
17500	23.076923076922\\
17600	52.1739130434726\\
17700	43.75\\
17800	0\\
17900	0\\
18000	38.4615384615317\\
18100	23.076923076922\\
18200	8.33333333332848\\
18300	12.5\\
18400	0\\
18500	21.0526315789466\\
18600	23.076923076922\\
18700	53.3333333333285\\
18800	22.222222222219\\
18900	25\\
19000	35.2941176470631\\
19100	26.470588235301\\
19200	20.8333333333285\\
19300	19.0476190476184\\
19400	18.1818181818235\\
19500	22.222222222219\\
19600	22.222222222219\\
19700	27.5862068965507\\
19800	22.7272727272793\\
19900	21.4285714285652\\
20000	29.0322580645152\\
20100	16.6666666666715\\
20200	29.0322580645152\\
20300	12.5\\
20400	21.7391304347839\\
20500	34.4827586206957\\
20600	38.2352941176505\\
20700	31.8181818181765\\
20800	33.3333333333285\\
20900	19.2307692307659\\
21000	38.8888888888905\\
21100	30.7692307692341\\
21200	31.4285714285652\\
21300	24.1379310344782\\
21400	16.6666666666715\\
21500	32.1428571428551\\
21600	26.0869565217436\\
21700	28.5714285714348\\
21800	25.8064516129089\\
21900	32.1428571428551\\
22000	32.1428571428551\\
22100	47.0588235294126\\
22200	14.2857142857101\\
22300	33.3333333333285\\
22400	20\\
22500	24\\
22600	17.8571428571449\\
22700	15\\
22800	20.6896551724203\\
22900	12\\
23000	18.1818181818235\\
23100	32.5\\
23200	41.0256410256407\\
23300	6.25\\
23400	30.555555555562\\
23500	31.0344827586232\\
23600	15.3846153846098\\
23700	12.5\\
23800	31.0344827586232\\
23900	39.2857142857101\\
24000	28.125\\
24100	37.5\\
24200	23.076923076922\\
24300	16.6666666666715\\
24400	15.7894736842136\\
24500	28.125\\
24600	38.8888888888905\\
24700	21.4285714285652\\
24800	31.25\\
24900	18.5185185185255\\
25000	37.9310344827536\\
25100	34.375\\
25200	36.6666666666715\\
25300	23.076923076922\\
25400	18.5185185185255\\
25500	24.2424242424313\\
25600	33.3333333333285\\
25700	25.925925925927\\
25800	14.8148148148175\\
25900	36.8421052631602\\
26000	21.4285714285652\\
26100	43.3333333333285\\
26200	32.3529411764757\\
26300	34.6153846153902\\
26400	19.0476190476184\\
26500	38.2352941176505\\
26600	19.2307692307659\\
26700	12.5\\
26800	25.925925925927\\
26900	35.2941176470631\\
27000	26.923076923078\\
27100	27.2727272727207\\
27200	30.7692307692341\\
27300	22.222222222219\\
27400	45.9459459459467\\
27500	38.4615384615317\\
27600	34.4827586206957\\
27700	32.1428571428551\\
27800	19.2307692307659\\
27900	34.375\\
28000	43.75\\
28100	37.5\\
28200	17.6470588235243\\
28300	29.0322580645152\\
28400	40\\
28500	20.5882352941117\\
28600	31.25\\
28700	20.8333333333285\\
28800	33.3333333333285\\
28900	20\\
29000	19.0476190476184\\
29100	37.9310344827536\\
29200	34.6153846153902\\
29300	15\\
29400	35.4838709677424\\
29500	17.6470588235243\\
29600	29.0322580645152\\
29700	46.4285714285652\\
29800	18.5185185185255\\
29900	31.0344827586232\\
30000	28.125\\
30100	15\\
30200	33.3333333333285\\
30300	30.7692307692341\\
30400	29.629629629635\\
30500	36\\
30600	28.5714285714348\\
30700	18.5185185185255\\
30800	47.3684210526262\\
30900	68.75\\
31000	52.9411764705874\\
31100	56.4102564102504\\
31200	50\\
31300	66\\
31400	71.1538461538439\\
31500	60.4166666666715\\
31600	63.1578947368398\\
31700	50\\
31800	66.6666666666715\\
31900	40.7407407407445\\
32000	63.0434782608645\\
32100	60\\
32200	60\\
32300	52.6315789473738\\
32400	57.1428571428551\\
32500	71.4285714285652\\
32600	65\\
32700	63.6363636363676\\
32800	38.7096774193487\\
32900	62.222222222219\\
33000	62.7906976744125\\
33100	56.6666666666715\\
33200	54.838709677424\\
33300	55.8139534883667\\
33400	60.4651162790688\\
33500	47.8260869565274\\
33600	61.1111111111095\\
33700	57.5\\
33800	61.9047619047633\\
33900	61.3636363636324\\
34000	71.4285714285652\\
34100	62.7906976744125\\
34200	70.2702702702663\\
34300	65\\
34400	60.9756097561039\\
34500	52.5\\
34600	47.5\\
34700	68.0851063829759\\
34800	55.263157894733\\
34900	56.25\\
35000	57.4468085106346\\
35100	51.4285714285652\\
35200	59.5238095238165\\
35300	63.6363636363676\\
35400	58.9743589743593\\
35500	59.0909090909117\\
35600	64.7058823529369\\
35700	47.3684210526262\\
35800	55.8823529411748\\
35900	60\\
36000	64.2857142857101\\
36100	51.515151515152\\
36200	68.0851063829759\\
36300	62.5\\
36400	68.5185185185255\\
36500	56.25\\
36600	58.5365853658586\\
36700	63.6363636363676\\
36800	44.444444444438\\
36900	60.4651162790688\\
37000	45.161290322576\\
37100	58.6956521739194\\
37200	71.4285714285652\\
37300	54.0540540540533\\
37400	59.375\\
37500	55.555555555562\\
37600	56.7567567567603\\
37700	44.736842105267\\
37800	56.4102564102504\\
37900	72.0930232558167\\
38000	46.875\\
38100	57.1428571428551\\
38200	68.9655172413768\\
38300	46.875\\
38400	73.9130434782564\\
38500	60\\
38600	48.5714285714348\\
38700	60.4651162790688\\
38800	38.2352941176505\\
38900	50\\
39000	69.5652173913113\\
39100	40\\
39200	54.5454545454559\\
39300	58.3333333333285\\
39400	40\\
39500	68.4210526315728\\
39600	61.2903225806513\\
39700	58.5365853658586\\
39800	53.125\\
39900	57.8947368421068\\
40000	62.222222222219\\
40100	57.777777777781\\
40200	61.1111111111095\\
40300	58.3333333333285\\
40400	47.6190476190532\\
40500	45.161290322576\\
40600	57.8947368421068\\
40700	59.4594594594528\\
40800	53.658536585368\\
40900	60.5263157894806\\
41000	66.6666666666715\\
41100	59.0909090909117\\
41200	62.5\\
41300	61.5384615384683\\
41400	64.7058823529369\\
41500	64.5161290322576\\
41600	68\\
41700	66.6666666666715\\
41800	53.125\\
41900	70.7317073170707\\
42000	61.9047619047633\\
42100	62.222222222219\\
42200	61.5384615384683\\
42300	67.3076923076878\\
42400	54.1666666666715\\
42500	62.1621621621598\\
42600	51.6129032258032\\
42700	60\\
42800	62.5\\
42900	60\\
43000	61.5384615384683\\
43100	54.2857142857101\\
43200	51.515151515152\\
43300	57.1428571428551\\
43400	51.4285714285652\\
43500	62.1621621621598\\
43600	61.9047619047633\\
43700	66.6666666666715\\
43800	60.9756097561039\\
43900	54.0540540540533\\
44000	67.6470588235243\\
44100	57.1428571428551\\
44200	61.1111111111095\\
44300	54.838709677424\\
44400	60\\
44500	57.5757575757598\\
44600	73.9130434782564\\
44700	52.777777777781\\
44800	57.5\\
44900	60\\
45000	64.7058823529369\\
45100	58.8235294117621\\
45200	53.5714285714348\\
45300	42.8571428571449\\
45400	50\\
45500	50\\
45600	48.3870967741968\\
45700	63.1578947368398\\
45800	62.5\\
45900	63.1578947368398\\
46000	71.1538461538439\\
46100	48.5714285714348\\
46200	65.2173913043516\\
46300	51.724137931029\\
46400	56.4102564102504\\
46500	56.6666666666715\\
46600	72.9729729729734\\
46700	52.9411764705874\\
46800	63.4146341463347\\
46900	64.2857142857101\\
47000	64\\
47100	54.2857142857101\\
47200	61.5384615384683\\
47300	65.7894736842136\\
47400	55.8823529411748\\
47500	43.3333333333285\\
47700	70.9677419354848\\
47800	65.1162790697708\\
47900	57.1428571428551\\
48000	54.5454545454559\\
48100	60\\
48200	71.7391304347839\\
48300	60.4166666666715\\
48400	55.555555555562\\
48500	54.0540540540533\\
48600	58.9743589743593\\
48700	52.1739130434726\\
48800	53.125\\
48900	42.8571428571449\\
49000	65.625\\
49100	64.7058823529369\\
49200	56.25\\
49300	55.8823529411748\\
49400	48.5714285714348\\
49500	38.0952380952367\\
49600	52.777777777781\\
49700	63.8888888888905\\
49800	70\\
49900	69.6969696969754\\
50000	48.3870967741968\\
50100	50\\
50200	51.2195121951227\\
50300	55.555555555562\\
50400	64.2857142857101\\
50500	72\\
50600	63.8888888888905\\
50700	36.6666666666715\\
50800	74.4186046511604\\
50900	55.555555555562\\
51000	60\\
51100	53.3333333333285\\
51200	60.5263157894806\\
51300	52\\
51400	62.7906976744125\\
51500	58.6206896551739\\
51600	60\\
51700	61.2903225806513\\
51800	50\\
51900	68.8888888888905\\
52000	55.8823529411748\\
52100	48.3870967741968\\
52200	53.3333333333285\\
52300	60.9756097561039\\
52400	40\\
52500	62.5\\
52600	56.0975609756133\\
52700	58.9743589743593\\
52800	53.8461538461561\\
52900	57.1428571428551\\
53000	50\\
53100	47.0588235294126\\
53200	69.0476190476184\\
53400	50\\
53500	71.4285714285652\\
53600	53.3333333333285\\
53700	59.5238095238165\\
53800	60\\
53900	67.3913043478242\\
54000	66.6666666666715\\
54100	56.6666666666715\\
54200	56.7567567567603\\
54300	63.8888888888905\\
54400	54.5454545454559\\
54500	64.444444444438\\
54600	62.7906976744125\\
54700	56.7567567567603\\
54800	61.1111111111095\\
54900	44.444444444438\\
55000	65\\
55100	63.4146341463347\\
55200	60.6060606060637\\
55300	55.263157894733\\
55400	66.6666666666715\\
55600	66.6666666666715\\
55700	55.555555555562\\
55800	66.6666666666715\\
55900	53.8461538461561\\
56000	68.4210526315728\\
56100	50\\
56200	67.5675675675739\\
56300	43.4782608695677\\
56400	62.1621621621598\\
56500	66.6666666666715\\
56600	62.5\\
56700	52.9411764705874\\
56800	56.4102564102504\\
56900	55.555555555562\\
57000	45.8333333333285\\
57100	76\\
57200	61.5384615384683\\
57300	68.8888888888905\\
57400	60.7142857142899\\
57500	60.9756097561039\\
57600	52.777777777781\\
57700	64.7058823529369\\
57800	51.6129032258032\\
57900	57.5\\
58000	55.555555555562\\
58100	40.7407407407445\\
58200	56.25\\
58300	33.3333333333285\\
58400	51.515151515152\\
58500	62.5\\
58600	61.2903225806513\\
58700	68.2926829268254\\
58800	71.4285714285652\\
58900	51.6129032258032\\
59000	59.375\\
59100	41.1764705882379\\
59200	53.5714285714348\\
59300	46.6666666666715\\
59400	64.7058823529369\\
59500	40.9090909090883\\
59600	67.8571428571449\\
59700	54.838709677424\\
59800	41.3793103448261\\
59900	45\\
60000	51.724137931029\\
60100	77.5\\
60200	60.7142857142899\\
60300	63.6363636363676\\
60400	46.4285714285652\\
60500	54.838709677424\\
60600	61.2903225806513\\
60700	52.9411764705874\\
60800	56.6666666666715\\
60900	51.6129032258032\\
61000	48\\
61100	57.5757575757598\\
61200	44.8275862068986\\
61300	53.5714285714348\\
61400	59.375\\
61500	51.6129032258032\\
61600	50\\
61700	45.8333333333285\\
61800	57.1428571428551\\
61900	59.375\\
62000	60.6060606060637\\
62100	54.2857142857101\\
62200	50\\
62300	38.0952380952367\\
62400	43.4782608695677\\
62500	47.3684210526262\\
62600	53.5714285714348\\
62700	51.724137931029\\
62800	54.5454545454559\\
62900	48.3870967741968\\
63000	45.4545454545441\\
63100	66.6666666666715\\
63200	58.9743589743593\\
63300	43.4782608695677\\
63400	52.9411764705874\\
63500	63.8888888888905\\
63600	40.9090909090883\\
63700	62.1621621621598\\
63800	57.1428571428551\\
63900	50\\
64000	60.7142857142899\\
64100	53.5714285714348\\
64200	51.724137931029\\
64300	41.1764705882379\\
64400	42.8571428571449\\
64500	39.1304347826081\\
64600	61.5384615384683\\
64700	38.0952380952367\\
64800	46.875\\
64900	54.5454545454559\\
65000	31.8181818181765\\
65100	66.6666666666715\\
65200	30\\
65300	45.4545454545441\\
65400	33.3333333333285\\
65500	58.8235294117621\\
65600	48.148148148146\\
65700	63.6363636363676\\
65800	68\\
65900	47.8260869565274\\
66000	50\\
66100	40.7407407407445\\
66200	50\\
66300	45.4545454545441\\
66400	44.8275862068986\\
66500	61.2903225806513\\
66600	40\\
66700	28.5714285714348\\
66800	41.1764705882379\\
66900	48\\
67000	59.2592592592555\\
67100	26.6666666666715\\
67200	64.2857142857101\\
67300	46.1538461538439\\
67400	38.8888888888905\\
67500	68.5714285714348\\
67600	44.444444444438\\
67700	22.222222222219\\
67800	45.4545454545441\\
67900	39.1304347826081\\
68000	27.777777777781\\
68100	52.3809523809468\\
68200	47.8260869565274\\
68300	44.8275862068986\\
68400	62.5\\
68500	45.8333333333285\\
68600	36.8421052631602\\
68700	60\\
68800	52.3809523809468\\
68900	52\\
69000	48.3870967741968\\
69100	50\\
69200	45.4545454545441\\
69300	33.3333333333285\\
69400	40\\
69500	54.2857142857101\\
69600	46.1538461538439\\
69700	40.7407407407445\\
69800	38.4615384615317\\
69900	51.724137931029\\
70000	33.3333333333285\\
70100	44.444444444438\\
70200	43.3333333333285\\
70300	39.2857142857101\\
70400	52.3809523809468\\
70500	45.4545454545441\\
70600	39.1304347826081\\
70700	55.1724137931014\\
70800	37.5\\
70900	59.2592592592555\\
71000	41.6666666666715\\
71100	52.1739130434726\\
71200	56.6666666666715\\
71300	40.7407407407445\\
71400	40\\
71500	45.8333333333285\\
71600	65.625\\
71700	57.8947368421068\\
71800	39.2857142857101\\
71900	60\\
72000	37.5\\
72100	46.6666666666715\\
72200	35\\
72300	39.1304347826081\\
72400	41.1764705882379\\
72500	40.7407407407445\\
72600	34.7826086956484\\
72700	50\\
72800	26.6666666666715\\
72900	69.444444444438\\
73000	50\\
73100	48.148148148146\\
73200	46.1538461538439\\
73300	50\\
73400	28.5714285714348\\
73500	40\\
73600	52.9411764705874\\
73700	41.3793103448261\\
73800	31.8181818181765\\
73900	44.444444444438\\
74000	23.529411764699\\
74100	45\\
74200	35\\
74300	36.3636363636324\\
74400	56.5217391304323\\
74500	56.6666666666715\\
74600	50\\
74700	47.0588235294126\\
74800	33.3333333333285\\
74900	36\\
75000	42.8571428571449\\
75100	42.8571428571449\\
75200	35.2941176470631\\
75300	33.3333333333285\\
75400	41.6666666666715\\
75500	30.4347826086887\\
75600	52.1739130434726\\
75700	44\\
75800	44.8275862068986\\
75900	26.6666666666715\\
76000	36.8421052631602\\
76100	45.8333333333285\\
76200	56.6666666666715\\
76300	8.33333333332848\\
76400	38.8888888888905\\
76500	55.555555555562\\
76600	26.923076923078\\
76700	37.0370370370365\\
76800	26.3157894736796\\
76900	47.8260869565274\\
77000	46.6666666666715\\
77100	47.8260869565274\\
77200	44\\
77300	52\\
77400	45.4545454545441\\
77500	36.3636363636324\\
77600	23.529411764699\\
77700	23.529411764699\\
77800	42.8571428571449\\
77900	21.4285714285652\\
78000	44.8275862068986\\
78100	29.4117647058883\\
78200	47.3684210526262\\
78300	33.3333333333285\\
78400	34.7826086956484\\
78500	35\\
78600	35.2941176470631\\
78700	33.3333333333285\\
78800	44.8275862068986\\
78900	17.6470588235243\\
79000	39.1304347826081\\
79100	42.8571428571449\\
79200	34.4827586206957\\
79300	37.0370370370365\\
79400	40\\
79500	21.4285714285652\\
79600	26.923076923078\\
79700	18.75\\
79800	52\\
79900	56.25\\
80000	21.4285714285652\\
80100	25\\
80200	33.3333333333285\\
80300	34.7826086956484\\
80400	26.6666666666715\\
80500	50\\
80600	31.25\\
80700	41.1764705882379\\
80800	30.7692307692341\\
80900	50\\
81000	18.1818181818235\\
81100	21.4285714285652\\
81200	45\\
81300	35.7142857142899\\
81400	38.0952380952367\\
81500	31.5789473684272\\
81600	46.1538461538439\\
81700	26.3157894736796\\
81800	47.0588235294126\\
81900	55.555555555562\\
82000	42.1052631578932\\
82100	59.375\\
82200	20\\
82300	47.0588235294126\\
82400	40\\
82500	50\\
82600	21.4285714285652\\
82700	23.529411764699\\
82800	14.2857142857101\\
82900	38.0952380952367\\
83000	31.5789473684272\\
83100	47.6190476190532\\
83200	40\\
83300	33.3333333333285\\
83400	25\\
83500	40\\
83600	20\\
83700	26.6666666666715\\
83800	25\\
83900	40\\
84000	35.2941176470631\\
84100	56.25\\
84200	30.7692307692341\\
84300	40.9090909090883\\
84400	33.3333333333285\\
84500	41.1764705882379\\
84600	9.09090909091174\\
84700	29.4117647058883\\
84800	12.5\\
84900	30.7692307692341\\
85000	21.4285714285652\\
85100	27.2727272727207\\
85200	21.4285714285652\\
85300	22.222222222219\\
85400	38.8888888888905\\
85500	35.7142857142899\\
85600	21.4285714285652\\
85700	39.1304347826081\\
85800	48.148148148146\\
85900	50\\
86000	27.777777777781\\
86100	31.5789473684272\\
86200	30.7692307692341\\
86300	42.8571428571449\\
86400	42.8571428571449\\
86500	23.8095238095266\\
86600	34.7826086956484\\
86700	28.5714285714348\\
86800	40\\
86900	30\\
87000	37.5\\
87100	57.8947368421068\\
87200	41.6666666666715\\
87300	33.3333333333285\\
87400	21.0526315789466\\
87500	50\\
87600	35.7142857142899\\
87700	37.5\\
87800	31.25\\
87900	28.5714285714348\\
88000	11.1111111111095\\
88100	28.5714285714348\\
88200	26.6666666666715\\
88300	35.2941176470631\\
88400	20\\
88500	33.3333333333285\\
88600	31.5789473684272\\
88700	28.5714285714348\\
88800	40\\
88900	30.7692307692341\\
89000	40.9090909090883\\
89100	20\\
89200	23.8095238095266\\
89300	14.2857142857101\\
89400	36.8421052631602\\
89500	15.3846153846098\\
89600	25\\
89700	12.5\\
89800	22.222222222219\\
89900	41.1764705882379\\
90000	7.14285714285506\\
90100	55\\
90200	37.5\\
90300	43.75\\
90400	36.3636363636324\\
90500	23.529411764699\\
90600	22.222222222219\\
90700	42.8571428571449\\
90800	46.1538461538439\\
90900	28.5714285714348\\
91000	10\\
91100	22.222222222219\\
91200	18.1818181818235\\
91300	31.25\\
91400	40\\
91500	0\\
91600	41.6666666666715\\
91700	44.444444444438\\
91800	10\\
91900	23.529411764699\\
92000	33.3333333333285\\
92100	23.529411764699\\
92200	41.6666666666715\\
92300	30.7692307692341\\
92400	18.1818181818235\\
92500	27.2727272727207\\
92600	21.0526315789466\\
92700	23.529411764699\\
92800	13.3333333333285\\
92900	0\\
93000	17.6470588235243\\
93100	40\\
93200	21.4285714285652\\
93300	16.6666666666715\\
93400	20\\
93500	30\\
93600	14.2857142857101\\
93700	14.2857142857101\\
93800	22.7272727272793\\
93900	23.076923076922\\
94000	26.6666666666715\\
94100	26.6666666666715\\
94200	11.1111111111095\\
94300	36.3636363636324\\
94400	44.444444444438\\
94500	38.4615384615317\\
94600	21.0526315789466\\
94700	31.25\\
94800	33.3333333333285\\
94900	18.75\\
95000	42.8571428571449\\
95100	27.2727272727207\\
95200	58.3333333333285\\
95300	27.777777777781\\
95400	9.09090909091174\\
95500	35.7142857142899\\
95600	25\\
95700	12.5\\
95800	28.5714285714348\\
95900	16.6666666666715\\
96000	0\\
96100	0\\
96200	30\\
96300	14.2857142857101\\
96400	0\\
96500	16.6666666666715\\
96600	21.4285714285652\\
96700	26.3157894736796\\
96800	23.076923076922\\
96900	0\\
97000	25\\
97100	25\\
97200	28.5714285714348\\
97300	0\\
97400	17.6470588235243\\
97500	14.2857142857101\\
97600	21.7391304347839\\
97700	46.1538461538439\\
97800	25\\
97900	0\\
98000	36.8421052631602\\
98100	33.3333333333285\\
98200	16.6666666666715\\
98300	20\\
98400	20\\
98500	30.7692307692341\\
98600	60\\
98700	0\\
98800	0\\
98900	15.3846153846098\\
99000	10\\
99100	37.5\\
99200	30\\
99300	7.69230769231217\\
99400	20\\
99500	14.2857142857101\\
99600	33.3333333333285\\
99700	23.076923076922\\
99800	11.1111111111095\\
99900	11.1111111111095\\
};
\addplot [color=red, line width=1.0pt, forget plot]
  table[row sep=crcr]{%
100	20\\
30900	29.9837925445754\\
31000	60\\
55100	59.9760138586571\\
100000	20.0879491849191\\
};
\end{axis}
%\end{tikzpicture}%

%% file: img/SimUncetainty.tex
% This file was created by matlab2tikz.
%
%The latest updates can be retrieved from
%  http://www.mathworks.com/matlabcentral/fileexchange/22022-matlab2tikz-matlab2tikz
%where you can also make suggestions and rate matlab2tikz.
%
%\begin{tikzpicture}

\begin{axis}[%
width=0.4 \textwidth,
height=0.9cm,
at={(0cm,1.6cm)},
scale only axis,
xmin=0,
xmax=100000,
ymin=0,
ymax=6,
y label style={at={(axis description cs:0.07,0.5)}},
ylabel={$u_G$ in \%},
axis background/.style={fill=white},
axis x line*=bottom,
axis y line*=left
]
\addplot [color=blue, forget plot]
  table[row sep=crcr]{%
100	4.7058823529369\\
200	2.38959609327139\\
300	1.55654151193448\\
400	1.20144665072439\\
500	1.0077134915482\\
600	0.891283867065795\\
700	0.810897610193933\\
800	0.756208691251231\\
900	0.727202845577267\\
1000	0.693886361681507\\
1100	0.672503989349934\\
1300	0.650331001583254\\
1500	0.63480505431653\\
1700	0.62616499971773\\
1800	0.619327106789569\\
1900	0.620051658930606\\
2000	0.628390338970348\\
2100	0.62385322956834\\
2200	0.627328463131562\\
2300	0.622103731438983\\
2700	0.631330914053251\\
3000	0.623483235292952\\
3100	0.626068626748747\\
3300	0.620688922994304\\
3400	0.624946327108773\\
3500	0.619396551643149\\
4100	0.617531658208463\\
4200	0.621599287420395\\
4300	0.620777218151488\\
4400	0.613509298651479\\
4700	0.620259893796174\\
4900	0.615402909912518\\
5100	0.616508389866794\\
5200	0.616999360601767\\
5300	0.621199081215309\\
5600	0.615671671213931\\
5700	0.619240387488389\\
5800	0.619027639346314\\
5900	0.622723616106668\\
6100	0.623440059978748\\
6200	0.616389190181508\\
6300	0.616910816504969\\
6400	0.622098516061669\\
6500	0.623082419653656\\
6600	0.618989425842301\\
6700	0.62076217662252\\
6900	0.616379316357779\\
7000	0.621697574315476\\
7100	0.617019317607628\\
7300	0.615744269874995\\
7500	0.612949680144084\\
7600	0.615299160490395\\
7900	0.615182693029055\\
8100	0.621640308410861\\
8300	0.613559904144495\\
8400	0.621494722698117\\
8600	0.621073018963216\\
8800	0.626400240624207\\
9100	0.623961626872187\\
9200	0.621560419094749\\
9300	0.623651884438004\\
9500	0.618905129696941\\
9700	0.635836482950253\\
9800	0.636324819686706\\
9900	0.632662502786843\\
10000	0.635961612279061\\
10100	0.629408625056385\\
10200	0.631504016637336\\
10600	0.621899858786492\\
10700	0.62488054469577\\
11000	0.615220056875842\\
11300	0.618124244210776\\
11400	0.616294289793586\\
11500	0.625521879672306\\
11600	0.626614146123757\\
11800	0.619554909571889\\
12000	0.630221290281042\\
12200	0.629058104081196\\
12300	0.6262989884126\\
12400	0.627194733635406\\
12500	0.623946216059267\\
12600	0.631359563165461\\
12700	0.62898709368892\\
12800	0.634189674761728\\
13000	0.631004687544191\\
13100	0.622850127663696\\
13200	0.619777724030428\\
13300	0.629169419320533\\
13400	0.627363698178669\\
13600	0.616384299748461\\
14300	0.620864188458654\\
14400	0.625078341297922\\
14500	0.619493086502189\\
14600	0.623073304159334\\
14700	0.619942067976808\\
15100	0.6226949942793\\
15200	0.615847296838183\\
15300	0.622262686563772\\
15400	0.625152852589963\\
15800	0.619392450666055\\
15900	0.620100294123404\\
16000	0.616799040173646\\
16200	0.624966365765431\\
16300	0.623271820659284\\
16500	0.626904160744743\\
16600	0.638670382555574\\
16900	0.62790818636131\\
17000	0.626428226489224\\
17100	0.635261148228892\\
17200	0.624964973379974\\
17400	0.618661252912716\\
17500	0.621481209906051\\
17600	0.633468262894894\\
17700	0.635558231369941\\
17800	0.628125154107693\\
17900	0.625609186012298\\
18200	0.627091264803312\\
18400	0.616945333778858\\
18500	0.626019170726067\\
18700	0.629683930295869\\
18800	0.634717908425955\\
18900	0.632472255529137\\
19000	0.635818017937709\\
19100	5.79710144927958\\
19200	2.7979729322833\\
19300	1.84932441794081\\
19400	1.41544945324131\\
19500	1.18886515712074\\
19600	1.04453909084259\\
19700	0.952038550822181\\
19800	0.873485411386355\\
19900	0.828721167446929\\
20000	0.800781632191502\\
20100	0.769386428015423\\
20200	0.756202145566931\\
20300	0.7365717026114\\
20400	0.720670966926264\\
20500	0.716436601855094\\
20600	0.719617187816766\\
20700	0.706780556633021\\
20900	0.699471643136349\\
21000	0.708885728832684\\
21100	0.703689393340028\\
21200	0.710939842174412\\
21300	0.708996317625861\\
21400	0.701305433656671\\
21500	0.700379646877991\\
21600	0.693604850675911\\
21700	0.694479347410379\\
21800	0.698793356932583\\
22000	0.69819832414214\\
22100	0.705386029585497\\
22200	0.694944902061252\\
22300	0.705318205218646\\
22500	0.69554684124887\\
22600	0.695968659347272\\
22700	0.68672875755874\\
22800	0.690388381772209\\
23000	0.683433430429432\\
23100	0.701106941807666\\
23200	0.7139762507868\\
23300	0.69524761888897\\
23400	0.705556917077047\\
23500	0.704871083537\\
23700	0.695009057541029\\
24000	0.701901750027901\\
24100	0.705782341727172\\
24200	0.701335432939231\\
24300	0.688349559583003\\
24400	0.679882580254343\\
24600	0.700341710384237\\
24700	0.699641999744927\\
24800	0.704028198757442\\
25000	0.701550982674235\\
25100	0.705510133353528\\
25200	0.706079374765977\\
25400	0.69935023067228\\
25500	0.70504215080291\\
25600	0.698326800440555\\
25800	0.695780310503324\\
25900	0.708477774765925\\
26100	0.70635060708446\\
26200	0.711756381031591\\
26300	0.705864757299423\\
26400	0.695301934130839\\
26500	0.703119142592186\\
26800	0.693581318017095\\
26900	0.70177159285231\\
27000	0.69828978717851\\
27100	0.704215597710572\\
27300	0.698271203800687\\
27400	0.709195271774661\\
27500	0.70392405858729\\
27700	0.70214911547373\\
27800	0.698576632377808\\
28100	0.709573034648201\\
28200	0.693226421877625\\
28600	0.709670803698828\\
28700	0.701813134932308\\
28800	0.69954261679959\\
28900	0.689392077591037\\
29000	0.68298357007734\\
29100	0.68749757684418\\
29200	0.687418947709375\\
29300	0.680344317137497\\
29400	0.687815758239594\\
29500	0.677184149957611\\
29600	0.685355135530699\\
30000	0.698254047703813\\
30100	0.688432207127335\\
30200	0.692900328474934\\
30500	0.689431321152369\\
30600	0.699740615746123\\
30700	0.697961796424352\\
30800	0.710209383163601\\
30900	0.733106511630467\\
31000	0.732559694661177\\
31100	0.738898301016889\\
31200	0.739768219602411\\
31400	0.779862568728277\\
31500	0.789778327205568\\
31600	0.782172797786188\\
31700	0.767344034960843\\
31800	0.779659575142432\\
31900	0.758184136851924\\
32000	0.769266919887741\\
32100	0.776665160155972\\
32200	0.767559457250172\\
32300	0.764921159658115\\
32400	0.748589005583199\\
32500	0.755779131752206\\
32600	0.758599051070632\\
32700	0.766650806763209\\
32800	0.754127944892389\\
32900	0.764536287519149\\
33000	0.769904666478396\\
33100	0.75516650712234\\
33200	0.745401983323973\\
33400	0.762053983009537\\
33500	0.739536270441022\\
33600	0.740265154483495\\
33800	0.754009073963971\\
33900	0.762982738306164\\
34100	0.772016926654032\\
34200	0.766917150467634\\
34600	0.769085197345703\\
34700	0.779546498946729\\
34800	0.774241169958259\\
34900	0.761325264247716\\
35000	0.773286426119739\\
35100	0.764963569032261\\
35300	0.774759221982094\\
35400	0.772009754626197\\
35500	0.777344294372597\\
35600	0.792426859639818\\
35700	0.784222374772071\\
35800	0.771864479916985\\
35900	0.763870349794161\\
36000	0.767898551013786\\
36100	0.75793570530368\\
36300	0.782252473509288\\
36400	0.801221787856775\\
36500	0.806976258041686\\
36600	0.800229692700668\\
36700	0.799732030005543\\
36800	0.78676079113211\\
36900	0.787512523442274\\
37000	0.769903424719814\\
37200	0.790388111665379\\
37300	0.781116468890104\\
37400	0.766543262026971\\
37600	0.758539872680558\\
37800	0.758816813395242\\
37900	0.76535749064351\\
38000	0.754567493408103\\
38100	0.760526179190492\\
38200	0.746679942327319\\
38300	0.740302235120907\\
38400	0.754842520007514\\
38500	0.757861687947297\\
38600	0.753081417802605\\
38700	0.760791136504849\\
38900	0.74863172924961\\
39000	0.761569119829801\\
39100	0.741920834654593\\
39200	0.738014066548203\\
39300	0.739079315346316\\
39400	0.725154149666196\\
39500	0.731708263221662\\
39600	0.727485425988561\\
39700	0.737602044144296\\
39800	0.733340637743822\\
39900	0.73814975717687\\
40100	0.76257639969117\\
40200	0.758153036003932\\
40400	0.719150409000576\\
40500	0.717842739031767\\
40800	0.740391425759299\\
41000	0.74766468044254\\
41100	0.757905818143627\\
41200	0.748885506545776\\
41300	0.751752211508574\\
41500	0.739150893816259\\
41600	0.75963869394036\\
41700	0.773416888769134\\
41800	0.760698974729166\\
42000	0.767942196980584\\
42100	0.775605602044379\\
42200	0.772670075341011\\
42300	0.790162793346099\\
42400	0.761612534188316\\
42500	0.758843881703797\\
42700	0.738726912619313\\
42800	0.745146063549328\\
42900	0.764556963869836\\
43000	0.764038598790648\\
43200	0.750244266891968\\
43300	0.757094165543094\\
43400	0.752488827056368\\
43500	0.751745465415297\\
43600	0.758286325610243\\
43700	0.781372134326375\\
43800	0.780196727253497\\
44300	0.744630115674227\\
44400	0.749810068344232\\
44500	0.744079923373647\\
44600	0.757894896887592\\
44700	0.754527655226411\\
44800	0.757613764872076\\
44900	0.767332947507384\\
45200	0.739247184785199\\
45300	0.720057316997554\\
45500	0.69688231634791\\
45600	0.700654947082512\\
45800	0.724214335452416\\
45900	0.730967862895341\\
46000	0.755760117142927\\
46100	0.751458527331124\\
46200	0.73171119739709\\
46300	0.724851073711761\\
46400	0.732809585490031\\
46500	0.727005622553406\\
46600	0.73182495860965\\
46700	0.731566885559005\\
46800	0.74084898197907\\
47000	0.718694600218441\\
47100	0.722677214420401\\
47300	0.736380498434301\\
47400	0.735094376694178\\
47500	0.728749218644225\\
47600	0.730514345676056\\
47700	0.726570212340448\\
47800	0.739598443295108\\
48000	0.735723147794488\\
48300	0.775700599318952\\
48400	0.755236052253167\\
48600	0.755677135457518\\
48700	0.734830011657323\\
48800	0.731211145102861\\
48900	0.723160792433191\\
49200	0.722967686597258\\
49400	0.727357149255113\\
49500	0.711271794891218\\
49800	0.733137569317478\\
50000	0.727129964958294\\
50100	0.720064293243922\\
50200	0.731689500491484\\
50300	0.722217125192401\\
50400	0.716331608462497\\
50500	0.740858432152891\\
50600	0.741294748499058\\
50700	0.733474880486028\\
50800	0.745131482181023\\
50900	0.744619675650029\\
51000	0.749801826255862\\
51100	0.739944295841269\\
51200	0.743335440129158\\
51300	0.728336434345692\\
51400	0.741014711631578\\
51500	0.731919916215702\\
51600	0.732981157620088\\
51700	0.728460822618217\\
51800	0.731625610191259\\
51900	0.746425911711412\\
52100	0.736015715054236\\
52200	0.729451981722377\\
52300	0.739166932587977\\
52400	0.725219704312622\\
52500	0.723815019082394\\
52600	0.734679186090943\\
52700	0.740570208858117\\
52800	0.727590443086228\\
53100	0.7321092831844\\
53300	0.751061479284544\\
53400	0.738166285009356\\
53500	0.74748602200998\\
53600	0.738184823858319\\
53700	0.747500792436767\\
53800	0.731446661855443\\
53900	0.747676437124028\\
54000	0.746598072379129\\
54100	0.737509864149615\\
54300	0.740667653430137\\
54400	0.73704924917547\\
54500	0.750800001551397\\
54600	0.758972921379609\\
54800	0.753672905892017\\
54900	0.738766557333292\\
55100	0.751651912098168\\
55200	0.745494086164399\\
55300	0.747686406190041\\
55400	0.759362062410219\\
55500	0.755664341748343\\
55600	0.736172947319574\\
55700	0.725601554499008\\
55800	0.745708261878463\\
55900	0.73144378780853\\
56000	0.736658499197802\\
56100	0.732615994988009\\
56200	0.736222592677223\\
56300	0.720411417642026\\
56400	0.72664749858086\\
56700	0.725958718307083\\
56800	0.733685236715246\\
57000	0.712366220745025\\
57100	0.737585199531168\\
57200	0.742861012622598\\
57300	0.755480737352627\\
57400	0.741494469926693\\
57500	0.748732436390128\\
57700	0.743622488647816\\
57800	0.736600669682957\\
57900	0.743464393017348\\
58000	0.743322956579505\\
58100	0.730989102667081\\
58200	0.728257671144092\\
58300	0.715760243139812\\
58400	0.717800349812023\\
58500	0.728554118555621\\
58600	0.725066868792055\\
58800	0.74550161602383\\
59000	0.733673384733265\\
59100	0.710853893280728\\
59300	0.707746122410754\\
59400	0.71284537082829\\
59500	0.701730414279155\\
59600	0.700704862872954\\
59800	0.703380092149018\\
59900	0.692248165447381\\
60000	0.694642602422391\\
60100	0.710086478546145\\
60200	0.707090634547058\\
60300	0.711067621479742\\
60400	0.707839347567642\\
60600	0.710119830866461\\
60700	0.693628012522822\\
60900	0.700681923248339\\
61000	0.696247637970373\\
61100	0.702623144039535\\
61400	0.705381630847114\\
61600	0.707401816442143\\
61700	0.700104773015482\\
61800	0.699460610834649\\
62200	0.715747197507881\\
62300	0.702659370479523\\
62500	0.685069155384554\\
62700	0.691309222514974\\
62800	0.69876828615088\\
62900	0.702115079620853\\
63000	0.693704892139067\\
63100	0.6969765376125\\
63200	0.710687890314148\\
63300	0.701346631496563\\
63500	0.715475210759905\\
63600	0.703692989802221\\
63700	0.713478874225984\\
63800	0.718603013447137\\
63900	0.71871064111474\\
64100	0.709819536466966\\
64200	0.708138325149775\\
64300	0.692172342955018\\
64400	0.685067468250054\\
64500	0.682056347824982\\
64600	0.683261979269446\\
64700	0.678381951380288\\
64800	0.687467564130202\\
65000	0.679113029400469\\
65100	0.692805541169946\\
65200	0.684368888832978\\
65400	0.676206890799222\\
65500	0.708857710895245\\
65600	0.704908205385436\\
65700	0.70937009548652\\
66000	0.691077354320441\\
66100	0.69134270970244\\
66200	0.678639684687369\\
66300	0.676046545122517\\
66600	0.693500843568472\\
66700	0.694399559055455\\
66800	0.682051539755776\\
67000	0.684459359574248\\
67100	0.672429986210773\\
67200	0.67818185921351\\
67300	0.665581770503195\\
67400	0.661755820125109\\
67500	0.677872695843689\\
67600	0.68121903813153\\
67700	0.663376169541152\\
67900	0.666445047783782\\
68000	0.662401849913294\\
68200	0.665044458553893\\
68400	0.683731897457619\\
68500	0.682207824866055\\
68600	0.675301792245591\\
68700	0.682721091812709\\
68800	0.67797547148075\\
68900	0.678985040533007\\
69000	0.686757660965668\\
69100	0.698854367627064\\
69200	0.691260420469916\\
69300	0.677429834599025\\
69400	0.68437359337986\\
69500	0.695757055655122\\
69700	0.693360071250936\\
69800	0.676673730107723\\
69900	0.682619657920441\\
70000	0.674470064521302\\
70300	0.688088311610045\\
70500	0.678586974201608\\
70600	0.677151136886096\\
70700	0.682989055567305\\
70800	0.67247613510699\\
70900	0.677069494253374\\
71100	0.676058542667306\\
71200	0.683308858162491\\
71400	0.684637288111844\\
71500	0.682894938057871\\
71600	0.690991682538879\\
71700	0.681850213251892\\
71800	0.685445509690908\\
71900	0.678868018716457\\
72000	0.678513083752478\\
72100	0.66804849926848\\
72200	0.665810431266436\\
72300	0.667449491549633\\
72400	0.662055664448417\\
72500	0.669036922612577\\
72600	0.669903343296028\\
72700	0.676230013734312\\
72800	0.666363733951584\\
72900	0.682688782297191\\
73000	0.67680408799788\\
73200	0.681993546386366\\
73400	0.674859315098729\\
73500	0.676606099688797\\
73600	0.66887084247719\\
73700	0.676574408571469\\
73800	0.674486398100271\\
74000	0.662771743911435\\
74300	0.662783312000101\\
74400	0.66514488386747\\
74700	0.692956046434119\\
74800	0.685654526241706\\
74900	0.684837192631676\\
75200	0.668124875577632\\
75300	0.673718559279223\\
75700	0.675771747773979\\
75800	0.681921593160951\\
75900	0.670561191960587\\
76000	0.666588468826376\\
76100	0.669159019205836\\
76200	0.677945009854739\\
76300	0.664303112906055\\
76400	0.660798583528958\\
76600	0.67254059764673\\
76700	0.677119102721917\\
76800	0.671499024523655\\
76900	0.671774090049439\\
77000	0.663071538918302\\
77400	0.671321148096467\\
77500	0.670513130695326\\
77700	0.659735050983727\\
77800	0.668359383838833\\
77900	0.659456047156709\\
78000	0.669260977650993\\
78100	0.663405687126215\\
78200	0.661217709450284\\
78300	0.668389839716838\\
78400	0.669411425769795\\
78700	0.662043379459647\\
78800	0.67127293725207\\
78900	0.66490410111146\\
79100	0.665946898967377\\
79200	0.674305354579701\\
79300	0.678476815839531\\
79500	0.659206888580229\\
79600	0.665726524297497\\
79700	0.659681063363678\\
79800	0.664984373404877\\
80100	0.645727013703436\\
80300	0.65538476324582\\
80400	0.650916852595401\\
80500	0.652875792351551\\
80800	0.644144581470755\\
80900	0.649846284708474\\
81000	0.642636554155615\\
81100	0.640385665232316\\
81200	0.644886764435796\\
81300	0.642061096485122\\
81400	0.647205187691725\\
81500	0.64900865573145\\
81600	0.644087744876742\\
81800	0.646493544321856\\
81900	0.656991566589568\\
82000	0.656392089033034\\
82100	0.670228503848193\\
82200	0.656496447918471\\
82300	0.653874373005237\\
82500	0.653906892985106\\
82600	0.648763343357132\\
82900	0.64901382959215\\
83100	0.653531532734632\\
83200	0.644306648217025\\
83400	0.644792507315287\\
83900	0.637134799486375\\
84100	0.639965740716434\\
84200	0.637378025392536\\
84400	0.652369535760954\\
84500	0.650783566583414\\
84600	0.643326052027987\\
84700	0.643994706682861\\
84800	0.635285016192938\\
85200	0.63158144953195\\
85400	0.639617583379732\\
85600	0.637030308760586\\
85800	0.656177626093267\\
86100	0.655053476104513\\
86200	0.648560424029711\\
86300	0.65215222563711\\
86400	0.647461301268777\\
86700	0.658100252097938\\
86800	0.658308613696136\\
86900	0.647802106803283\\
87000	0.646311576114385\\
87100	0.648333210585406\\
87200	0.655183208742528\\
87300	0.64760612178361\\
87400	0.649311637607752\\
87500	0.641213601993513\\
87700	0.639959035441279\\
88700	0.642523681410239\\
88800	0.63622781081358\\
88900	0.63459796445386\\
89000	0.642541528432048\\
89100	0.636240936451941\\
89200	0.642765816461178\\
89300	0.640481947004446\\
89400	0.643921676310129\\
89500	0.640316038290621\\
89600	0.644833848680719\\
89700	0.635897810017923\\
89900	0.641057667467976\\
90000	0.63920925410639\\
90100	0.643992533849087\\
90300	0.643059194466332\\
90400	0.649016963681788\\
90600	0.648758819836075\\
90700	0.652303065086016\\
91100	0.631005528848618\\
91200	0.628724764770595\\
91400	0.638495022925781\\
91500	0.62828923210327\\
91600	0.627694357041037\\
92000	0.646442508863402\\
92100	0.646336663019611\\
92300	0.638204812130425\\
92500	0.630981198119116\\
92600	0.6367128482525\\
92800	0.638705567587749\\
92900	0.629430785367731\\
93100	0.639663633803139\\
93200	0.638170008911402\\
93300	0.641140351988724\\
93400	0.640294181896024\\
93500	0.644817237785901\\
93600	0.64200934994733\\
93700	0.632831613111193\\
93800	0.641187399887713\\
94000	0.638158807734726\\
94200	0.641060062582255\\
94300	0.647488677408546\\
94400	0.638853925629519\\
94500	0.636551522358786\\
94600	0.6409423014411\\
94700	0.64117220985645\\
94800	0.637257644368219\\
94900	0.638406435537036\\
95000	0.644418594427407\\
95200	0.635380801264546\\
95300	0.639031536484254\\
95400	0.634663179749623\\
95600	0.636287347864709\\
95900	0.640020451290184\\
96000	0.628406136616832\\
96100	0.62192485286505\\
96200	0.621018563630059\\
96300	0.624218967597699\\
96400	0.619823549131979\\
96600	0.624494019954\\
96700	0.631776887268643\\
96800	0.631282592265052\\
96900	0.625977343748673\\
97000	0.629916416000924\\
97300	0.625132210363518\\
97400	0.630270485038636\\
97500	0.631153592388728\\
97600	0.64092556250398\\
97800	0.634974792963476\\
97900	0.627698914438952\\
98100	0.638150882165064\\
98200	0.635018991204561\\
98300	0.635712337010773\\
98400	0.631211855215952\\
98500	0.630861323690624\\
98700	0.617786726492341\\
98800	0.615097940171836\\
98900	0.618810065891012\\
99000	0.618708302514278\\
99100	0.6244271438336\\
99400	0.623025372507982\\
99600	0.625785469266702\\
99700	0.626811095935409\\
99900	0.621340107099968\\
};
\end{axis}

\begin{axis}[%
width=0.4 \textwidth,
height=0.9cm,
at={(0cm,0.0cm)},
scale only axis,
scale only axis,
xmin=0,
xmax=100000,
x label style={at={(axis description cs:0.5,0.1)}},
xlabel={packets},
ymin=0,
ymax=20,
y label style={at={(axis description cs:0.07,0.5)}},
ylabel={$u_B$ in \%},
axis background/.style={fill=white},
axis x line*=bottom,
axis y line*=left
]
\addplot [color=blue, forget plot]
  table[row sep=crcr]{%
100	18.1818181818235\\
200	10.8344231572992\\
300	10.0881061924156\\
400	8.45027852711792\\
500	7.0334303182899\\
600	5.86669163557235\\
700	5.18349751424103\\
800	4.65909258559986\\
900	3.95209224382415\\
1000	3.77694446593523\\
1100	3.55954164701689\\
1200	3.29176814598031\\
1300	3.09690244664671\\
1400	2.95499969754019\\
1500	2.85333434073254\\
1600	2.72660065966193\\
1700	2.67045579875412\\
1800	2.67271522605733\\
1900	2.57164178254607\\
2000	2.36671073766774\\
2100	2.34640611555369\\
2200	2.23686192900641\\
2300	2.24088101419329\\
2400	2.15974819882831\\
2500	2.04756948820432\\
2600	1.99384459466091\\
2700	1.95689319790108\\
2800	1.94321838767792\\
3000	1.94630856707226\\
3100	1.90610624274996\\
3200	1.90678997382929\\
3300	1.92577504905057\\
3400	1.87937235501886\\
3500	1.91869719146052\\
3600	1.91818334224808\\
3700	1.89950492215576\\
3900	1.89299769462377\\
4000	1.88598650587664\\
4100	1.91570432104345\\
4200	1.87947486867779\\
4300	1.88266916235443\\
4400	1.9499056962668\\
4600	1.91616889317811\\
4700	1.89771692616341\\
4800	1.92663777180132\\
4900	1.94408048102923\\
5000	1.93175713061646\\
5200	1.93685814578203\\
5300	1.89788829359168\\
5400	1.91756058720057\\
5500	1.95462838334788\\
5600	1.95065953287121\\
5700	1.91905100946315\\
5800	1.91850340865494\\
5900	1.88191280365572\\
6100	1.86205515029724\\
6200	1.92065956318402\\
6300	1.91995845059864\\
6400	1.87435556080891\\
6500	1.86055804729403\\
6600	1.89198861843033\\
6700	1.87623898277525\\
6900	1.91726765382919\\
7000	1.872034071901\\
7100	1.911788055324\\
7200	1.91193220792047\\
7500	1.97222974813485\\
7600	1.95692724437686\\
7700	1.97198932753236\\
7800	2.00577139688539\\
8100	1.91161665515392\\
8300	1.98421578887792\\
8400	1.91122474004806\\
8600	1.90250926646695\\
8700	1.85929351783125\\
8900	1.82876674570434\\
9000	1.84512898781395\\
9100	1.83453264532727\\
9200	1.85041503699904\\
9300	1.83082457535784\\
9500	1.86970327254676\\
9600	1.78995991579723\\
9700	1.72374124717317\\
9800	1.70392140701006\\
9900	1.7155089036969\\
10000	1.68244638032047\\
10100	1.71789175133745\\
10200	1.69880457155523\\
10500	1.73249573775684\\
10600	1.7644245151605\\
10700	1.74704836342426\\
10800	1.77793781878427\\
10900	1.79042328200012\\
11000	1.84330595334177\\
11100	1.84134540232481\\
11200	1.86531727801776\\
11300	1.85251431573124\\
11400	1.87564588646637\\
11500	1.80296569092025\\
11600	1.78887544483587\\
11800	1.84254336130107\\
12000	1.75427138988744\\
12200	1.74567015115463\\
12300	1.76101475242467\\
12400	1.75168596701405\\
12500	1.77433687617304\\
12600	1.71805922991189\\
12700	1.72831431742816\\
12800	1.68631948773691\\
12900	1.67816169963044\\
13000	1.6921525508078\\
13100	1.74958342257014\\
13200	1.78029105467431\\
13300	1.71567950714962\\
13400	1.72615953229251\\
13500	1.76630374693195\\
13600	1.82032350543886\\
13800	1.84532879806648\\
13900	1.86900148712448\\
14000	1.84722345561022\\
14100	1.86204507308139\\
14200	1.83265277944156\\
14300	1.84018691563688\\
14400	1.80534749972867\\
14500	1.84893374826061\\
14600	1.82118709890347\\
14700	1.84666438520071\\
15100	1.82952836321783\\
15200	1.88942162215244\\
15300	1.83948596657137\\
15400	1.81291904284444\\
15500	1.82225389161613\\
15600	1.84765187252196\\
15900	1.85580846203084\\
16000	1.88755668929662\\
16100	1.84637196850963\\
16200	1.81894570823351\\
16300	1.82773240774986\\
16500	1.78857468519709\\
16600	1.69341613931465\\
17000	1.74382237053942\\
17100	1.67813838113216\\
17200	1.74376838511671\\
17400	1.80377253018378\\
17500	1.78958977328148\\
17600	1.70144037358114\\
17700	1.67733922463958\\
17800	1.72050705798029\\
17900	1.73804986150935\\
18200	1.72735435780487\\
18300	1.75964851345634\\
18400	1.81397964760254\\
18500	1.75137262143835\\
18600	1.7431247348577\\
18700	1.72086210128327\\
18800	1.67996589720133\\
18900	1.68664737982908\\
19000	1.65758530580206\\
19100	1.52657868983806\\
19200	1.47935372707434\\
19400	1.43066561965679\\
19600	1.34728596072819\\
19700	1.30757491409895\\
19800	1.30411396901764\\
20000	1.24144055024954\\
20100	1.23931911363616\\
20200	1.21127308030555\\
20400	1.21907562756678\\
20500	1.20184003756731\\
20600	1.17012036457891\\
20700	1.18507030195906\\
20900	1.17995999904815\\
21000	1.14577975396242\\
21100	1.15042588545475\\
21200	1.1251691874204\\
21300	1.12325240155042\\
21400	1.13758760147903\\
21500	1.13692358134722\\
21600	1.15273620594235\\
22000	1.13621861765569\\
22100	1.11674067324202\\
22200	1.1416355711699\\
22300	1.11491740170459\\
22500	1.13774207756796\\
22600	1.13705418832251\\
22700	1.16290475653659\\
22800	1.1549498153181\\
23000	1.17765300643805\\
23100	1.13098964255187\\
23200	1.09720648855728\\
23300	1.14041875759722\\
23400	1.1139343805844\\
23600	1.12303832828184\\
23700	1.13740211891127\\
24000	1.12076727462409\\
24100	1.11022241652245\\
24200	1.11998264222348\\
24400	1.1817332294886\\
24600	1.13014598663722\\
24700	1.13062874504249\\
24800	1.11838790719048\\
25000	1.12211523880251\\
25200	1.10851502676087\\
25400	1.12392940559948\\
25500	1.10975454488653\\
25600	1.12588378670625\\
25800	1.13388388554449\\
25900	1.10254011218785\\
26100	1.10506170972076\\
26200	1.09120735904435\\
26300	1.10364896591636\\
26400	1.13008155756688\\
26500	1.11172097235976\\
26800	1.13870218055672\\
26900	1.11877071594063\\
27000	1.12731334986165\\
27100	1.11254100235237\\
27300	1.12687290100439\\
27400	1.09994504999486\\
27600	1.11144365021028\\
27900	1.11281234085618\\
28100	1.09600205122842\\
28200	1.13608788565034\\
28600	1.09645023597113\\
28700	1.11433052089706\\
28800	1.12035971348814\\
29000	1.1692624084244\\
29100	1.16027456217853\\
29200	1.16279871005099\\
29300	1.1856698456395\\
29400	1.16714452278393\\
29500	1.20021476995316\\
29600	1.17914114885207\\
30000	1.14291296488955\\
30100	1.16809033222671\\
30200	1.1559430539055\\
30500	1.1646108279092\\
30600	1.1366629479453\\
30700	1.13937798507686\\
30800	1.1069328252488\\
30900	7.69230769231217\\
31000	4.68215045802935\\
31100	3.25716561246372\\
31200	2.57434682901658\\
31300	2.00140893252683\\
31400	1.64801211941813\\
31500	1.44262370678189\\
31600	1.34319685328228\\
31700	1.2915960142418\\
31800	1.19050891014922\\
31900	1.18501668462704\\
32000	1.11770806832646\\
32100	1.06919607974123\\
32300	1.04213770762726\\
32400	1.0557231760904\\
32600	1.01265843791771\\
32700	0.989650644754875\\
32800	1.00304015213624\\
33000	0.965703051449964\\
33200	0.999172997646383\\
33400	0.967093412284157\\
33500	1.00360891764285\\
33700	0.991050053024082\\
34100	0.944644504459575\\
34300	0.949375258176588\\
34800	0.935660069910227\\
34900	0.954746625327971\\
35000	0.936861286303611\\
35100	0.948966785013909\\
35500	0.929987003648421\\
35600	0.908904022435308\\
36100	0.958551117422758\\
36500	0.889223855236196\\
37100	0.924724020296708\\
37200	0.908907751640072\\
37600	0.956521924337721\\
37800	0.957829846796812\\
37900	0.948203901920351\\
38000	0.965416408915189\\
38100	0.956588296307018\\
38300	0.992105916287983\\
38400	0.968614863872062\\
38500	0.963792813432519\\
38600	0.971518075544736\\
38700	0.959207471314585\\
38900	0.979119564159191\\
39000	0.958367200408247\\
39100	0.990938215443748\\
39600	1.02493787663116\\
39700	1.00691827214905\\
39800	1.01509782084031\\
39900	1.00655478288536\\
40100	0.963265873360797\\
40200	0.968725998987793\\
40400	1.0403406094847\\
40500	1.04591533397615\\
40900	0.998371690249769\\
41200	0.985594612531713\\
41300	0.980003871809458\\
41500	1.00119515664119\\
41600	0.966341859748354\\
41700	0.943782874877797\\
41800	0.961658134197933\\
42200	0.93979377085634\\
42300	0.914450632481021\\
42400	0.954542562642018\\
42500	0.959161150050932\\
42700	0.994845945941051\\
42800	0.985118487777072\\
42900	0.953917169899796\\
43100	0.963387604118907\\
43200	0.975918449898018\\
43300	0.965061405484448\\
43500	0.974085950801964\\
43600	0.963584141471074\\
43700	0.928503647097386\\
43900	0.937800705229165\\
44100	0.961668219853891\\
44500	0.986895790629205\\
44600	0.964506762422388\\
44800	0.964724110468524\\
44900	0.949223019808414\\
45200	0.995476739451988\\
45300	1.03404123065411\\
45500	1.09435342879442\\
45600	1.09126244677464\\
45900	1.03035084399744\\
46000	0.983898897422478\\
46100	0.988290868554031\\
46200	1.02237173325557\\
46300	1.03609184504603\\
46400	1.02104800858069\\
46500	1.03229070361704\\
46800	1.00577717399574\\
47000	1.04881709891197\\
47200	1.0258920773922\\
47300	1.01535778802645\\
47600	1.02560086705489\\
47700	1.0334798777767\\
47800	1.00865636905655\\
48000	1.01416420853639\\
48300	0.943271063297288\\
48400	0.972912714321865\\
48600	0.970824815958622\\
48700	1.00691486796131\\
49200	1.03845915674174\\
49400	1.03168130667473\\
49500	1.06629972855444\\
49800	1.02430194185581\\
50000	1.03472984868858\\
50100	1.04937926710409\\
50200	1.02649212934193\\
50400	1.05820756935282\\
50500	1.0100523228175\\
50600	1.00748413785186\\
50700	1.02075782700558\\
50800	0.998551351964124\\
51200	0.997377418680117\\
51300	1.02512007136829\\
51400	1.0020190183277\\
51500	1.01869266352151\\
51800	1.02093374085962\\
51900	0.993729103924125\\
52100	1.01093786042475\\
52200	1.02369645578437\\
52300	1.00592173935729\\
52400	1.03258607555472\\
52500	1.0366825692472\\
52700	1.00503292233043\\
52800	1.02915504595148\\
53100	1.0212894720753\\
53300	0.985610936870216\\
53400	1.00716065519373\\
53500	0.99016901224968\\
53600	1.00600393525383\\
53700	0.989242047406151\\
53800	1.01800448182621\\
53900	0.988970860344125\\
54000	0.99005686910823\\
54100	1.00590826496773\\
54300	1.0004620555992\\
54400	1.00711372002843\\
54600	0.968281273264438\\
54800	0.974766808300046\\
54900	1.00029886815173\\
55100	0.978413042088505\\
55200	0.988582265912555\\
55300	0.984861309276312\\
55400	0.965224961342756\\
55500	0.970354880933883\\
55600	1.00397234629781\\
55700	1.02559700225538\\
55800	0.98999337614805\\
55900	1.01607432313904\\
56000	1.00735181692289\\
56100	1.01546307509125\\
56200	1.00939370134438\\
56300	1.04100817277504\\
56400	1.03030504719936\\
56700	1.03439834128949\\
56800	1.01967741549015\\
57000	1.0643859802949\\
57100	1.01475660472352\\
57200	1.00375111906033\\
57300	0.980177154400735\\
57400	1.00247367072734\\
57500	0.988850274909055\\
57700	0.995807018553023\\
57800	1.00826394125761\\
57900	0.995988355076406\\
58000	0.995864644239191\\
58100	1.01858332216216\\
58200	1.02491734255455\\
58300	1.05191680962162\\
58400	1.05012068826181\\
58500	1.02972857283021\\
58600	1.03696492136805\\
58800	0.997694798192242\\
59000	1.01757979468675\\
59100	1.06497105697053\\
59300	1.07815253736044\\
59400	1.06905635242583\\
59500	1.09646632123622\\
59700	1.09770743790432\\
59800	1.10008474357892\\
59900	1.13011730801372\\
60000	1.12741781516524\\
60100	1.09137683200242\\
60200	1.09773632831639\\
60300	1.08813845022814\\
60500	1.09178761618386\\
60600	1.08911613524833\\
60700	1.12986146057665\\
60900	1.11603459739126\\
61000	1.12814057944342\\
61100	1.11322195750836\\
61400	1.10648463737743\\
61600	1.10017030370363\\
61700	1.11756270227488\\
61800	1.11997086685733\\
62200	1.07833442771516\\
62500	1.15757603447128\\
62700	1.14731808965735\\
62900	1.12015564447211\\
63000	1.14138079034456\\
63100	1.13366726983804\\
63200	1.0993371548102\\
63300	1.11996599155827\\
63500	1.08396544247807\\
63600	1.10959532285051\\
63800	1.07265138429648\\
63900	1.07021290529519\\
64200	1.09171129226161\\
64300	1.13220847574121\\
64500	1.16868163671461\\
64600	1.1699651722447\\
64700	1.18845555373991\\
64800	1.16603503779334\\
65000	1.19499507834553\\
65100	1.15783540502889\\
65300	1.19480234030925\\
65400	1.21022039075615\\
65500	1.12083661583893\\
65600	1.12590575453942\\
65700	1.11138211068464\\
66100	1.15365093816945\\
66200	1.19161941172206\\
66300	1.20380037624273\\
66700	1.15376177901635\\
66800	1.188179795834\\
67000	1.18466157093644\\
67100	1.22338597929047\\
67200	1.20906745416869\\
67300	1.25334621513321\\
67400	1.27335898725141\\
67500	1.22393474246201\\
67600	1.21319481389946\\
67700	1.27314131827734\\
67900	1.2715231324255\\
68000	1.28946822122089\\
68200	1.28714989169384\\
68400	1.22297609975794\\
68500	1.22351359813183\\
68600	1.24299152138701\\
68700	1.21797906835855\\
68800	1.23048239838681\\
68900	1.22617256666126\\
69000	1.20049026700144\\
69100	1.16223436438304\\
69200	1.17818961865851\\
69300	1.21750450048421\\
69400	1.19694080657791\\
69500	1.16276367814862\\
69700	1.1633671758027\\
69800	1.21131641788816\\
69900	1.19538353767712\\
70000	1.22182379134756\\
70300	1.18268992948288\\
70500	1.21075382261188\\
70600	1.21672335767653\\
70700	1.19988339419069\\
70800	1.23339340934763\\
70900	1.22114737951779\\
71100	1.22637979904539\\
71200	1.20427520737576\\
71400	1.19731563552341\\
71500	1.20149447137373\\
71600	1.17672356103139\\
71700	1.20155880249513\\
71800	1.19078916995204\\
71900	1.21035860982374\\
72000	1.21269442740595\\
72100	1.24882988810714\\
72200	1.26134709083999\\
72300	1.26028115896042\\
72400	1.28361396824766\\
72500	1.26321607327554\\
72600	1.26188574990374\\
72700	1.24118347324838\\
72800	1.27463578485185\\
72900	1.22121079305361\\
73000	1.23711840156466\\
73200	1.21717502742831\\
73400	1.23749343804957\\
73500	1.23213599105657\\
73600	1.25847712183895\\
73700	1.23452448220633\\
73800	1.24105638188485\\
74000	1.28555140996468\\
74200	1.3003702340211\\
74400	1.29161881828622\\
74700	1.19307715872128\\
74800	1.20871032650757\\
74900	1.20762302385992\\
75200	1.25969599226664\\
75300	1.24321301071905\\
75700	1.23784523690119\\
75800	1.21743104849884\\
75900	1.25312403963471\\
76000	1.26912243565312\\
76100	1.26295422053954\\
76200	1.23440641709021\\
76300	1.28068589592294\\
76400	1.29757930425694\\
76600	1.25986444299633\\
76700	1.24335409574269\\
76800	1.2605081814545\\
76900	1.25956082025368\\
77000	1.29125454378664\\
77200	1.27319503223407\\
77300	1.26242841275234\\
77500	1.26759189044242\\
77700	1.31021404551575\\
77800	1.28127574203245\\
77900	1.31517351759248\\
78000	1.28125884420297\\
78200	1.31237879979017\\
78300	1.2871947063104\\
78400	1.28244401916163\\
78500	1.29075061972253\\
78600	1.31076280177513\\
78700	1.31113875831943\\
78800	1.27794452302624\\
78900	1.29936041572364\\
79100	1.29562796591199\\
79200	1.26518694088736\\
79300	1.24781041609822\\
79500	1.31458798461244\\
79600	1.29320023946639\\
79700	1.31726631741913\\
79800	1.29965150018688\\
80000	1.3530562868109\\
80100	1.38980471099785\\
80300	1.36090584340855\\
80400	1.38251228473382\\
80500	1.3777404588036\\
80800	1.42871023507905\\
80900	1.4076506497513\\
81000	1.44496547269227\\
81100	1.46304045416764\\
81200	1.44715889509826\\
81300	1.46501096386055\\
81400	1.44362208309758\\
81500	1.43560083242483\\
81600	1.45993343564624\\
81700	1.44972540681192\\
81800	1.45134442327253\\
81900	1.40185283616302\\
82000	1.39933842272148\\
82100	1.33646577470063\\
82200	1.38410447989008\\
82300	1.39353186280641\\
82500	1.39147065968427\\
82600	1.41489337255189\\
82700	1.4206947699422\\
82800	1.44121719598479\\
82900	1.42331490651122\\
83100	1.40344405181531\\
83200	1.44630440282344\\
83400	1.45077508236864\\
83900	1.51445413130568\\
84100	1.50816374049464\\
84200	1.52549725466815\\
84400	1.44853687871364\\
84500	1.45030000517727\\
84600	1.48417121001694\\
84700	1.48157078474469\\
84800	1.53022385721852\\
85200	1.58219322915829\\
85400	1.54273448117601\\
85600	1.55761205412273\\
85700	1.51071248832159\\
85800	1.45024608496169\\
85900	1.43616424241918\\
86100	1.42748504996416\\
86200	1.45258218697563\\
86300	1.43302038239199\\
86400	1.45230428825016\\
86700	1.3933000544348\\
86800	1.38707005143806\\
86900	1.43111705846968\\
87000	1.44014747056644\\
87100	1.43258941355452\\
87200	1.40106117195683\\
87300	1.43374265564489\\
87400	1.42703598870139\\
87500	1.46816557476996\\
87600	1.48386828560615\\
88500	1.50002258080349\\
88600	1.4843537544657\\
88700	1.49838224078121\\
88800	1.53412714654405\\
88900	1.54888517457584\\
89000	1.50911908039416\\
89100	1.54403411300154\\
89200	1.51073121929949\\
89300	1.52199839783134\\
89400	1.50328425400949\\
89500	1.52109761894098\\
89600	1.49688630813034\\
89700	1.54458370857174\\
89900	1.52036575759121\\
90000	1.5306140841858\\
90100	1.50501198656275\\
90300	1.50586116200429\\
90400	1.47293356509181\\
90600	1.4652660368738\\
90700	1.44383952581848\\
91000	1.52008615477826\\
91100	1.56021024911024\\
91200	1.58480255979521\\
91300	1.57558503383189\\
91400	1.54331431354512\\
91500	1.6072053098178\\
91600	1.62105695136415\\
92000	1.51347097894177\\
92100	1.50721642728604\\
92500	1.59506843089184\\
92600	1.5659660865349\\
92700	1.55301574191253\\
92800	1.55371198964713\\
92900	1.61065674771089\\
93000	1.59185625627288\\
93100	1.55713367299177\\
93200	1.56344343149976\\
93300	1.54482983065827\\
93400	1.54646436322946\\
93500	1.51852999404946\\
93600	1.52897287570522\\
93700	1.58086378526059\\
93800	1.5359134603641\\
93900	1.55049291689647\\
94100	1.55235151732631\\
94200	1.53524090872088\\
94300	1.49766049579193\\
94400	1.53936238421011\\
94500	1.55359656363726\\
94600	1.53043958646595\\
94700	1.52785518327437\\
94800	1.54921582122915\\
94900	1.5443623947358\\
95000	1.51100894031697\\
95200	1.5600753849867\\
95300	1.5419191056717\\
95400	1.5680958799785\\
95500	1.5732159542531\\
95700	1.5585971402179\\
95800	1.56474848356447\\
95900	1.54595745938423\\
96000	1.61624256710638\\
96100	1.67003109600046\\
96200	1.69190016076027\\
96300	1.68309544361546\\
96400	1.72591070980707\\
96500	1.72792505775578\\
96600	1.71490072652523\\
96700	1.6678967144835\\
96800	1.66879495975445\\
96900	1.70518393396924\\
97000	1.68059045016707\\
97100	1.68720921248314\\
97200	1.67894840598456\\
97300	1.71463619831775\\
97400	1.68169546770514\\
97500	1.67407233902486\\
97600	1.60704002025886\\
97800	1.62752386579814\\
97900	1.67369884542131\\
98000	1.63297043510829\\
98100	1.60473382503551\\
98200	1.61882424484065\\
98300	1.61182583509071\\
98400	1.6385452964314\\
98500	1.64256895068684\\
98700	1.75148214760702\\
98800	1.79002826131182\\
98900	1.7774165890296\\
99000	1.78994859610975\\
99100	1.75396547402488\\
99200	1.76857937275781\\
99300	1.75839904758323\\
99400	1.77262118351064\\
99500	1.75425716891186\\
99700	1.74487055338977\\
99900	1.78940037662687\\
};
\end{axis}
%\end{tikzpicture}%